\documentclass{aa}
\usepackage[dvips]{graphicx}
\newcommand{\gtrsim}{\mathrel{\hbox{\rlap{\hbox{\lower4pt\hbox{$\sim$}}}\hbox{$>$}}}}
\newcommand{\lesssim}{\mathrel{\hbox{\rlap{\hbox{\lower4pt\hbox{$\sim$}}}\hbox{$<$}}}}

\def\CO{$^{12}$CO}

\def\METH{CH$_3$OH }
\def\MCN{\mbox{CH$_3$CN}}
\def\MCNI{$^{13}$CH$_3$CN}
\def\MCNII{CH${_3}^{13}$CN}

\def\nh{N$_2$H$^+$}
\def\cs{C$^{34}$S}
\def\HII{H{\sc ii} }

\def\kms{\mbox{km~s$^{-1}$}}

\def\cmq{cm$^{-2}$}

\def\Vlsr{$V_{\rm LSR}$}

\def\jykms{~Jy~beam$^{-1}$\,km~s$^{-1}$}
\def\jy{~Jy~beam$^{-1}$}
\def\mjy{~mJy~beam$^{-1}$}

\def\jduk{\mbox{2$_k$--1$_k$}}
\def\jdo{\mbox{12--11}}
\def\jcc{\mbox{5--4}}
\def\jdu{\mbox{2--1}}
\def\juz{\mbox{1--0}}
\def\jtd{\mbox{3--2}}
\def\jsc{\mbox{6--5}}

\begin{document}
\title{A detailed study of the rotating toroids in G31.41+0.31 and G24.78+0.08\thanks{Based on observations carried out with the IRAM
Plateau de Bure Interferometer.  IRAM is supported by INSU/CNRS
(France), MPG (Germany) and IGN (Spain).}} 
\author{M.\ T. Beltr\'an\inst{1} \and R.\ Cesaroni \inst{1} \and R.\ Neri \inst{2}
\and C.\ Codella \inst{3} \and R.\ S.\ Furuya \inst{4} \and L.\ Testi \inst{1} \and L.\ Olmi \inst{3}} 
\institute{
INAF, Osservatorio Astrofisico di Arcetri, Largo E. Fermi 5,
50125 Firenze, Italy
\and
IRAM, 300 Rue de la Piscine, F-38406 Saint Martin d'H\`eres, France 
\and
Istituto di Radioastronomia, INAF, Sezione di Firenze, Largo E. Fermi 5,
50125 Firenze, Italy
\and
Division of Physics, Mathematics, and Astronomy,
California Institute of Technology, MS 105-24, Pasadena, CA 91125, USA}

\offprints{M. T. Beltr\'an, \email{mbeltran@arcetri.astro.it}}
\date{Received date; accepted date}

\titlerunning{The rotating toroids in G31.41+0.31 and G24.78+0.08}
\authorrunning{Beltr\'an et al.}

\abstract{We present the results of high angular resolution millimeter
observations of gas and dust toward G31.41+0.31 and G24.78+0.08, two high-mass
star forming regions where four rotating massive toroids have been previously
detected by Beltr\'an et al.~(\cite{beltran04}). The \MCN\ (\jdo) emission of
the toroids in G31.41+0.31 and core A1 in G24.78+0.08 has been modeled assuming
that it arises from a disk-like structure seen edge-on, with a radial velocity
field. For G31.41+0.31 the model properly fits the data for a velocity $v_{\rm
rot}\simeq 1.7$~\kms\ at the outer radius $R_{\rm out}\simeq 13400$~AU and an
inner radius $R_{\rm inn}\simeq 1340$~AU, while for core A1 in G24.78+0.08 the best fit is obtained for $v_{\rm rot}\simeq 2.0$~\kms\ at $R_{\rm out}\simeq
7700$~AU and $R_{\rm inn}\simeq 2300$~AU. Unlike the rotating disks detected
around less luminous stars, these toroids are not undergoing Keplerian
rotation. From the modeling itself, however, it is not possible to distinguish
between constant rotation or constant angular velocity, since both velocity
fields suitably fit the data. The best fit models have been computed adopting a
temperature gradient of the type $T \propto R^{-3/4}$, with a temperature at
the outer radius $T_{\rm out}\simeq 100$~K for both cores. The $M_{\rm dyn}$
needed for equilibrium derived from the models is much smaller than the mass of
the cores, suggesting that such toroids are unstable and undergoing
gravitational collapse. The collapse is also supported by the \MCNII\ or \MCN\
line width measured in the cores, which increases toward the center of the
toroids. The estimates of $v_{\rm inf}$ and $\dot M_{\rm acc}$ are $\sim
2$~\kms\ and $\sim 3\times10^{-2}~M_\odot$~yr$^{-1}$ for G31.41+0.31, and $\sim
1.2$~\kms\ and $\sim 9\times10^{-3}~M_\odot$~yr$^{-1}$ for G24.78+0.08 A1. Such
large accretion rates could weaken the effect of stellar winds and radiation
pressure and allow further accretion on the star. The values of $T_{\rm rot}$
and $N_{\rm CH_3CN}$, derived by means of the RD method, for both G31.41+0.31
and the sum of cores A1 and A2 (core A of Codella et al.~\cite{codella97}) in
G24.78+0.08 are in the range 132--164~K and 2--$8\times10^{16}$~cm$^{-2}$. For
G31.41+0.31, the most plausible explanation for the apparent toroidal
morphology seen in the lower $K$ transitions of \MCN\ (\jdo) is
self-absorption, which is caused by the high optical depth and temperature
gradient in the core. 
}

\maketitle

\section{Introduction}

The formation of massive stars represents a puzzle from a theoretical point of
view. Unlike low-mass stars, they are believed to reach the
zero-age main sequence still deeply embedded in their parental cores: in
particular, Palla \& Stahler~(\cite{palla93}) predict that this occurs for
stellar masses in excess of 8~$M_\odot$. Once the star has ignited hydrogen
burning, further accretion should be inhibited by radiation pressure and
powerful stellar winds, with the consequence that stars more massive than
8~$M_\odot$ should not exist. Two formation scenarios have been proposed to
solve this paradox: non-spherical accretion (Nakano et al.~\cite{nakano95}; Jijina \& Adams~\cite{jijina96}) and merging of lower mass stars (Bonnell et
al.~\cite{bonnell98}). Discriminating between  these two models represents a
challenging observational goal.

In this context, the detection of disks would strongly favour the accretion
scenario, as naively one would expect that random encounters between merging
stars do not lead to axially symmetric structures since the disks associated
with the lower mass stars should be destroyed during the merging process. On
the contrary, conservation of angular momentum is bound to cause flattening and
rotation of the infalling material, thus producing disk-like bodies. Indeed,
circumstellar disks have been detected in low-mass stars and found to undergo
Keplerian rotation (Simon et al.~\cite{simon00}). Similar evidence has been
found in a few high-mass Young Stellar Objects (YSOs), but in most cases the
angular resolution was insufficient to assess the presence of a disk
unambiguously. In conclusion, only few bona fide examples are known (see
Cesaroni~\cite{cesa02}) and all of these are associated with moderately massive
stars (B1--B0). This is not sufficient to understand the role of disks in the
formation of even more massive stars and establish the relevance of accretion
to this process.

With this in mind, we have decided to perform a search for disks in high-mass
YSOs by means of high angular resolution observations. For this purpose, we
have selected  luminous ($L_{\rm bol}>10^4~L_\odot$) objects with typical signposts of
massive star formation such as water masers and ultracompact (UC) \HII regions.
The first objects selected for this project are G31.41+0.31 and G24.78+0.08.
The former is a well studied hot core located at 7.9~kpc (Cesaroni et
al.~\cite{cesa94}, \cite{cesa98}), where preliminary evidence of a rotating
massive disk oriented perpendicularly to a bipolar outflow has been reported in
Cesaroni et al.~(\cite{cesa94}) and Olmi et al.~(\cite{olmi96a}). The latter
one is a cluster of massive (proto)stars with a distance of 7.7~kpc, where
recently Furuya et al.~(\cite{furuya02}) have detected a pair of cores, each of
these associated with a compact bipolar outflow. On the basis of previous
experience with this type of objects (see e.g.\ Cesaroni et al.~\cite{cesa99}),
\MCN\ has been used as disk tracer. This is a low-abundance molecule which is
excited in very dense regions. Therefore, searching for disks requires not only
high angular resolution, but also great sensitivity given the faintness of the
lines observed. The first result derived from the analysis of the high angular
resolution data has been the detection of clear velocity gradients
perpendicular to the axis of the molecular outflows in the cores (see Fig.~2 in
Beltr\'an et al.~\cite{beltran04}, hereafter Paper I), which points out the
presence of rotating structures in the hot cores: one in G31.41+0.31 and three
in G24.78+0.08. In Paper~I, we referred to these rotating structures as
``disks''; however, their  sizes ($\sim 8000$--16000~AU in diameter) and masses
($\sim 80$--$500\,M_\odot$) make them remarkably different from the
geometrically thin circumstellar disks seen in low-mass YSOs, where
the mass of the disk is much less than that of the star. Thus, these structures
are more toroidal, and it is plausible to interpret such toroids as
``circumcluster'' disks, in the sense that they may host not just a single
star, but a whole cluster. The rotating toroidal structures detected in Paper~I
are unstable as suggested by the fact that the mass of the cores is much larger
than the dynamical mass needed for equilibrium. The main finding of the
preliminary results of such a study is that nonspherical accretion is a viable
mechanism to form high-mass stars.

In this article we present a full report on the results obtained in Paper~I,
and a complete analysis of the data. We thoroughly analyze all the molecular
lines detected toward the cores, the gas and dust distribution, and the
physical parameters. We study in detail the kinematics of the cores and model
the gas emission by assuming that it arises from a disk-like structure with an
internal radial velocity field, and by adopting power-law distributions for the
velocity and temperature of the emitting gas.

\section{Observations}

\subsection{BIMA observations}

\begin{table*}
\caption[] {Parameters of the BIMA observations}
\label{bima_setup}
\begin{tabular}{lcccrccc}
\hline
&&&\multicolumn{2}{c}{Synthesized beam} \\
\cline{4-5}
\multicolumn{1}{c}{Observation} &
\multicolumn{1}{c}{Configuration} &
\multicolumn{1}{c}{Frequency} &
\multicolumn{1}{c}{HPBW} &
\multicolumn{1}{c}{P.A.} &
\multicolumn{1}{c}{Bandwidth} &
\multicolumn{1}{c}{Spectral resolution}& 
\multicolumn{1}{c}{rms noise$^{\rm a}$}\\
& & \multicolumn{1}{c}{(GHz)} &
\multicolumn{1}{c}{(arcsec)} &
\multicolumn{1}{c}{(deg)} &
\multicolumn{1}{c}{(MHz)} &
\multicolumn{1}{c}{(\kms)}& 
\multicolumn{1}{c}{(mJy~beam$^{-1}$)}
\\
\hline
3.20~mm continuum &A, B &93.55 &$1.28\times0.94$ &0 &1200 &$-$ &1\\
\MCN\ (\jcc) &A, B  &91.98 &$0.83\times0.60$ &61 &25 &0.31 
 &50\\
3.16~mm continuum &C, D &94.75 &$19.8\times16.3$ &$-2$ &800 &$-$                   &3\\
N$_2$H$^+$ (1--0) &C, D &93.18 &$14.3\times13.0$ &16 &25 &0.31                    &100\\
C$^{34}$S (2--1) &C, D &96.41 &$12.6\times13.5$ &14 &25 &0.31                     &100\\
\METH\ (\mbox{2$_{k}$--1$_{k}$})  &C, D &96.74 &$14.0\times12.6$ &12 &25 &0.31&100 \\
\hline 

\end{tabular}

(a) For the molecular line observations the 1-$\sigma$ noise is per channel.

\end{table*}

Millimeter observations of G24.78+0.08 were carried out with the BIMA array\footnote{The BIMA array is operated by the Berkeley-Illinois-Maryland Association with support from the National Science Foundation.} in the A, B, C, and D configurations between 2002 September and 2003 February (see Table~\ref{bima_setup}). The digital correlator was configured to observe simultaneously the continuum emission and some molecular lines. Details of the observations are given in Table~\ref{bima_setup}. The phase center was located at $\alpha$(J2000) = 18$^{\rm h}$ 36$^{\rm m}$ $12\fs659$, $\delta$(J2000) = $-07\degr$ 12$'$ $10\farcs15$. The bandpass calibration was carried out on 3C345.
Amplitude and phase were calibrated by observations of 1743$-$048, whose flux density of 4.4~Jy at 3.20~mm (A and B configurations) and of 4.7~Jy at 3.16~mm (C and D configurations) was determined relative to Uranus, or Mars for the B configuration observations. The data were calibrated and analyzed using standard procedures in the MIRIAD software package. We subtracted the continuum from the line emission directly in the {\it(u, v)}-domain in order to reduce non linearity effects in the deconvolution, and thus any amplification of errors induced in this process.

\subsection{PdBI observations}

\begin{table*}
\caption[] {Frequency setups used for the molecular lines observed with the IRAM PdBI}
\label{freq_setup}
\begin{tabular}{lccccc}
\hline
\multicolumn{1}{c}{Line} &
\multicolumn{1}{c}{Center frequency} &
\multicolumn{1}{c}{Bandwidth} &
\multicolumn{2}{c}{Spectral resolution} &
\multicolumn{1}{c}{rms noise$^{\rm a}$}\\
&\multicolumn{1}{c}{(MHz)} &
\multicolumn{1}{c}{(MHz)} &
\multicolumn{1}{c}{(MHz)} &
\multicolumn{1}{c}{(\kms)}&
\multicolumn{1}{c}{(mJy~beam$^{-1}$)} \\
\hline
\MCN\ (\jcc),  \MCNII\ (\jcc) &91987.094 &80 &0.156 &0.5 &12  \\
\MCNII\ (\jcc) &91913.398 &320 &2.5 &8.1 &3 \\
\MCN\  (\jcc) $v_8=1$ &92353.438 &160 &0.625 &2.0 &7 \\
\MCN\  (\jcc) $v_8=1$ &92175.484 &320 &2.5 &8.1 &3 \\
\MCN\ (\jdo) &220747.266 &80 &0.156 &0.2  &50 \\
\MCN\ (\jdo),  \MCNII\ (\jdo) &220679.297 &80 &0.156 &0.2 &50 \\
\MCN\ (\jdo),  \MCNII\ (\jdo)  &220594.422 &160 &1.25 &1.7 &20\\
\MCN\ (\jdo),  \MCNII\ (\jdo),  $^{13}$CO (\jdu) &220475.812 &320 &2.5 &3.4 &16 \\
\hline
\end{tabular}

(a) For the molecular line observations the 1-$\sigma$ noise is per channel.

\end{table*}

Interferometric observations of G31.41+0.31 and G24.78+0.08 were carried out with the IRAM Plateau de Bure Interferometer
(PdBI) in the most extended configuration on 2003 March 16. By using the dual frequency capabilities of the PdBI we observed the continuum and molecular lines simultaneously at 3.3 and 1.4~mm. The receivers were tuned single-sideband at 3.3~mm and double-sideband at 1.4~mm. The correlator was centered at 92.0475~GHz in the lower sideband at 3.3~mm and at 220.5250~GHz in the upper sideband at 1.4~mm. The frequency setup of the correlator and the list of the observed molecular lines are shown in Table~\ref{freq_setup}. The eight units of the correlator were placed in such a way that a frequency range free of lines could be used to measure the continuum flux. The phase center was set to the position $\alpha$(J2000) = 18$^{\rm h}$ 47$^{\rm m}$ $34\fs330$, $\delta$(J2000) = $-01\degr$ 12$'$ $46\farcs50$ for G31.41+0.31, and to $\alpha$(J2000) = 18$^{\rm h}$ 36$^{\rm m}$ $12\fs661$, $\delta$(J2000) = $-07\degr$ 12$'$ $10\farcs15$ for G24.78+0.08. The bandpass of the receivers was calibrated by observations of the quasar 3C273. Amplitude and phase calibrations were achieved by monitoring 1741$-$038, whose flux densities of 4.63~Jy  and 2.99~Jy at 3.3~mm and 1.4~mm, respectively, were determined relative to MWC349. The data were calibrated and analyzed with the GILDAS software package developed at IRAM and Observatoire de Grenoble. We subtracted the continuum from the line emission directly in the {\it(u, v)}-domain. The synthesized CLEANed beams for maps made using natural weighting were $2\farcs7\times1\farcs2$ at P.A.\ $=-174\degr$ (3.3~mm) and $1\farcs1\times0\farcs5$ at P.A.\ $=-170\degr$ (1.4~mm) for G31.41+0.31, and  $3.0\farcs1\times1\farcs2$ at P.A.\ $=-177\degr$ (3.3~mm) and $1\farcs2\times0\farcs5$ at P.A.\ $=-174\degr$ (1.4~mm) for G24.78+0.08. Note  that the G24.78+0.04 BIMA and PdBI continuum data at 3~mm were merged in the ({\it u, v})-domain (see next section) and the synthesized CLEANed beam for the resulting map was  $1\farcs3\times0\farcs9$ at P.A.\ $=0\degr$. 	 

 One of the lines observed and clearly detected toward both regions is
$^{13}$CO (\jdu). However, in both cases the line profile exhibits a strong
lack of emission at the center, probably due to a missing flux problem since 
the interferometer might have filtered out part of the extended emission as
suggested by the single-dish spectra (G31.41+0.31: Cesaroni, private communication; G24.78+0.08: Cesaroni et al.~\cite{cesa03}). In addition,
although the line exhibits prominent wings, it has been impossible to
disentangle the core emission (at a systemic velocity) from the high-velocity
emission to derive the parameters of the outflowing material. Therefore, the
emission of $^{13}$CO (\jdu) has not been analyzed in the paper.

\subsection{Merging the BIMA and PdBI data for G24.78+0.08}

The BIMA (A and B configuration) and PdBI continuum data of G24.78+0.08 at 3~mm were merged
in the ({\it u, v})-domain. The projected baseline lengths range from $\sim3.8$ to 416~k$\lambda$ for the BIMA data, and from $\sim4.8$ to 127~k$\lambda$ for the PdBI data. The BIMA ({\it u, v})-data were first converted to
FITS format using a procedure available in the MIRIAD software package, and
then converted to GILDAS format using a procedure available in the GILDAS software
package. All the BIMA data were then merged together to create a single ({\it u,
v})-table using the task {\it uv\_merge} of GILDAS, and the
relative weights of the data were divided by 1$\times10^6$ so that the weight
density of the BIMA data matches that of the PdBI data. Finally, the  PdBI
({\it u, v})-data were merged with the already created BIMA ({\it u, v})-table.

\section{G31.41+0.31}

\subsection{Results}

\subsubsection{Continuum emission}

\begin{table*}
\caption[] {Positions, flux densities and sizes at millimeter wavelengths for the cores in G31.41+0.31 and G24.78+0.08}
\label{table_cont}
\begin{tabular}{lcccccccc}
\hline
&\multicolumn{2}{c}{Position$^{\rm a}$} &
&&&&\multicolumn{2}{c}{Source Diameter$^{\rm b}$} \\ 
 \cline{2-3} 
 \cline{8-9}
&\multicolumn{1}{c}{$\alpha({\rm J2000})$} &
\multicolumn{1}{c}{$\delta({\rm J2000})$} &
\multicolumn{1}{c}{$I^{\rm peak}_\nu$(3.3mm)} &
\multicolumn{1}{c}{$S_\nu$(3.3mm)} &
\multicolumn{1}{c}{$I^{\rm peak}_\nu$(1.4mm)} &
\multicolumn{1}{c}{$S_\nu$(1.4mm)} &
\multicolumn{1}{c}{$\theta_s$} &
\multicolumn{1}{c}{$\theta_s$} \\
\multicolumn{1}{c}{Core} &
\multicolumn{1}{c}{h m s}&
\multicolumn{1}{c}{$\degr$ $\arcmin$ $\arcsec$} &
\multicolumn{1}{c}{({Jy beam$^{-1}$})} & 
\multicolumn{1}{c}{(Jy)} &
\multicolumn{1}{c}{({Jy beam$^{-1}$})} & 
\multicolumn{1}{c}{(Jy)} &
\multicolumn{1}{c}{(arcsec)} &
\multicolumn{1}{c}{(AU)} \\
\hline
G31 &18 47 34.32 &$-01$ 12 45.9 & 0.252  &0.674  &1.3 &4.4 &1.1 &8700 \\
G24  A1 &18 36 12.57 &$-07$ 12 10.9 &~0.086$^{\rm c}$ &~0.211$^{\rm d}$ &0.191 &0.882 &1.2 &9200\\ 
G24  A2 &18 36 12.49 &$-07$ 12 10.0 &~0.086$^{\rm c}$ &~0.211$^{\rm d}$ &0.125 &0.574 &0.9 &7000\\
G24 B  &~18 36 12.66$^{\rm e}$ &~$-07$ 12 15.2$^{\rm e}$ &0.016 &0.040 &$<0.020^{\rm f}$   &$<0.020^{\rm f}$  &~1.5$^{\rm g}$ &11500$^{\rm g}$\\
G24 C  &18 36 13.12 &$-07$ 12 07.4 &0.005 &0.005 &0.056 &0.100 &0.9 &7000\\
\hline

\end{tabular}
   
  (a) Position of the 1.4~mm emission peak. \\
  (b) Deconvolved geometric mean of the major and minor axes of the 50\% 
  of the peak contour at 1.4~mm. \\
  (c) Peak intensity of core A (cores A1 and A2 together; see 
  Sect.~\ref{g24-cont}). \\ 
  (d) Flux density of core A (cores A1 and A2 together; see Sect.~\ref{g24-cont}). \\
  (e) Source not detected at 1.4~mm. Position corresponds to the 3.3~mm emission peak. \\
  (f) Upper limit for nondetected source is $3\sigma$. \\
  (g) Deconvolved geometric mean of the major and minor axes of the 50\% 
  of the peak contour at 3.3~mm.\\
\end{table*}

\begin{figure*}
\centerline{\includegraphics[angle=-90,width=11.5cm]{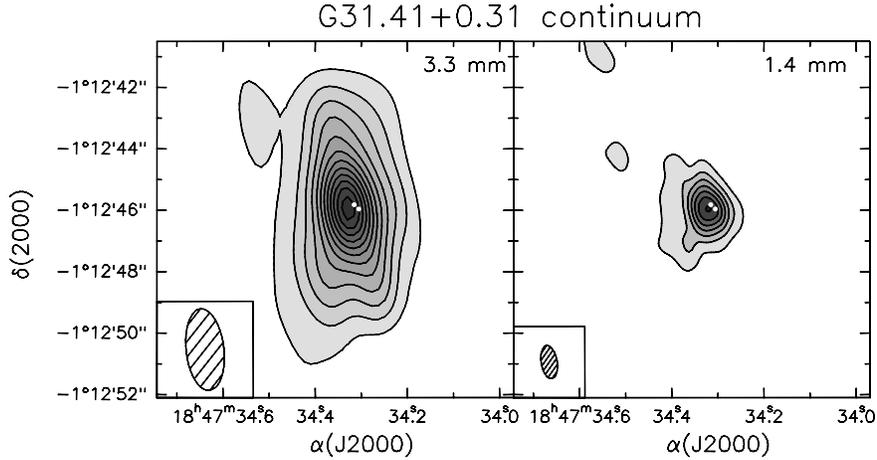}}
\caption{PdBI map of the 3.3~mm ({\it left panel}) and the 1.4~mm ({\it right panel}) continuum emission toward the core of G31.41+0.31. The contour levels range from 20 to 240\mjy\ in steps of 20\mjy\ for the 3.3~mm map, and from 100 to 1300\mjy\ in steps of  200\mjy\ for the 1.4~mm map. The synthesized beam is shown in the lower left-hand corner. The white dots mark the position of the 7~mm continuum emission peaks detected by Hofner (private communication).}
\label{g31_continuum}
\end{figure*}

In Fig.~\ref{g31_continuum} we show the PdBI maps of the 3.3 and 1.4~mm
continuum emission toward the hot core G31.41+0.31. The continuum dust emission
is marginally resolved at both wavelengths and shows a compact source plus an
extended envelope. With the BIMA interferometer, Wyrowski et al.\
(\cite{wyro01}) have mapped a similar structure at 1.4~mm. The position and fluxes
at 3.3 and 1.4~mm, and the deconvolved size of the source, measured as the
geometric mean of the major and minor axes of the 50\% of the peak contour at 1.4~mm, are given in  Table~\ref{table_cont}. The flux densities of 0.67~Jy at 3.3~mm and 4.4~Jy at 1.4~mm are
consistent with the values of 0.28~Jy at 3.4~mm and of 4.0~Jy at 1.3~mm measured by Maxia et al.\ (\cite{maxia01})  with the Owens Valley Radio Observatory (OVRO) interferometer. Hatchell et al.\ (\cite{hatchell00}) have
measured a flux of 4.9~Jy at 1.35~mm inside the $22''$ James Clerk
Maxwell Telescope (JCMT) beam. The fact that the continuum flux seen with a
single dish is not much higher than that measured with the interferometers
seems to indicate that the source is very centrally peaked, and that there is
no significant fraction of the total flux filtered out by the
interferometers. In Paper~I we derived a core mass of $490\,M_\odot$
from the millimeter continuum emission assuming a mass opacity of $\simeq0.02$~cm$^{-2}$g$^{-1}$ at 1.4~mm for a gas-to-dust ratio of 100 and a temperature of 230~K (Olmi et al.~\cite{olmi96a}).

\subsubsection{\MCN\ and \MCNII}
\label{g31-ch3cn}

\begin{figure}
\centerline{\includegraphics[angle=0,width=7.5cm]{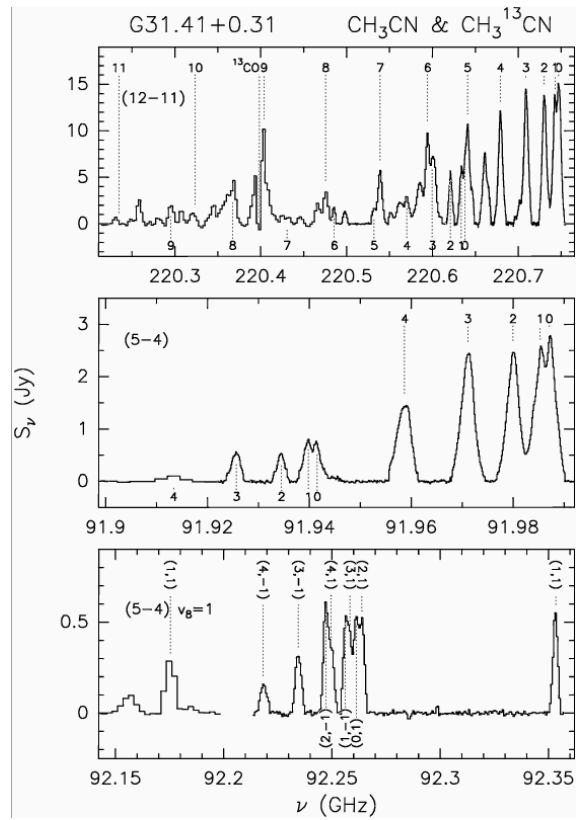}}
\caption{Methyl cyanide spectra integrated over the $3\sigma$ contour level area toward G31.41+0.31 as seen with the PdBI. We show in the top \MCN\ (\jdo), in the middle \MCN\ (\jcc), and in the bottom \MCN\ (\jcc) vibrationally excited ($v_8=1$). Different $K$ numbers ({\it top} and {\it middle panels}) are marked with dashed lines in the upper (lower) part of each spectra in the case of \MCN\ (\MCNII). Note that different $K$-components of a same transition have different spectral resolution because of the different spectral resolution of the correlator units (see Table~\ref{table_cont}).}
\label{g31-spectra}
\end{figure}

\begin{figure}
\centerline{\includegraphics[angle=0,width=7.5cm]{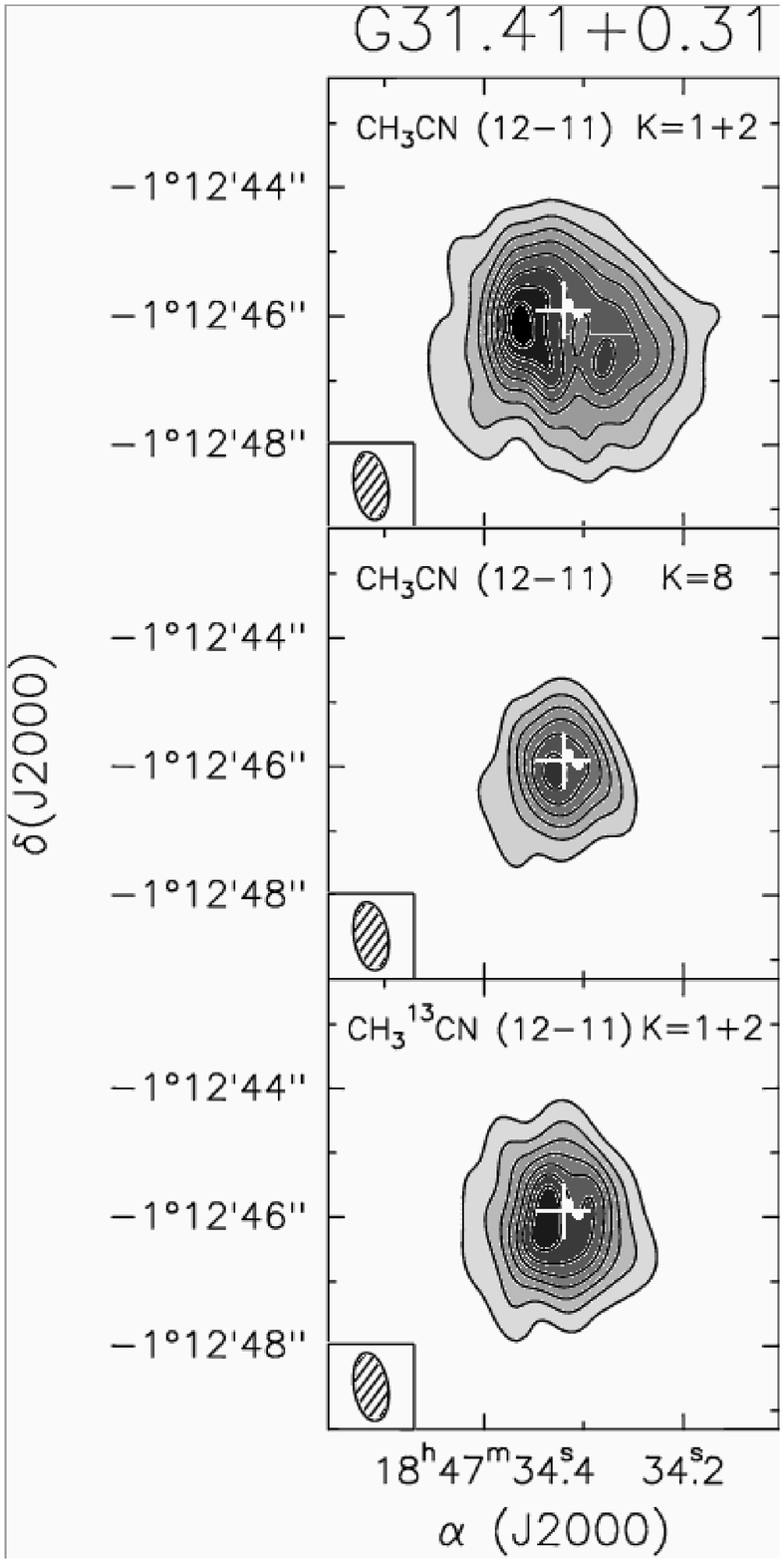}}
\caption{PdBI intensity maps of the \MCN\ (\jdo) emission averaged under the $K=1$ and 2 components ({\it top panel}), under the $K=8$ component ({\it middle panel}), and of the \MCNII\ (\jdo) emission averaged under the $K=1$ and 2 components ({\it bottom panel}) toward G31.41+0.31. The contour levels range 
from 0.1 to 0.94\jy\ ({\it top panel}), to 0.70\jy\ ({\it middle panel}), and to 0.82\jy\ ({\it bottom panel}), in steps of 
0.12\jy. The synthesized beam is shown in the lower left-hand corner.
The white cross marks the position of the 1.4~mm continuum emission peak, and the white dots the 7~mm continuum emission peaks detected by Hofner (private communication).}
\label{g31_ch3cn_1mm}
\end{figure}

\begin{figure}
\centerline{\includegraphics[angle=0,width=7.5cm]{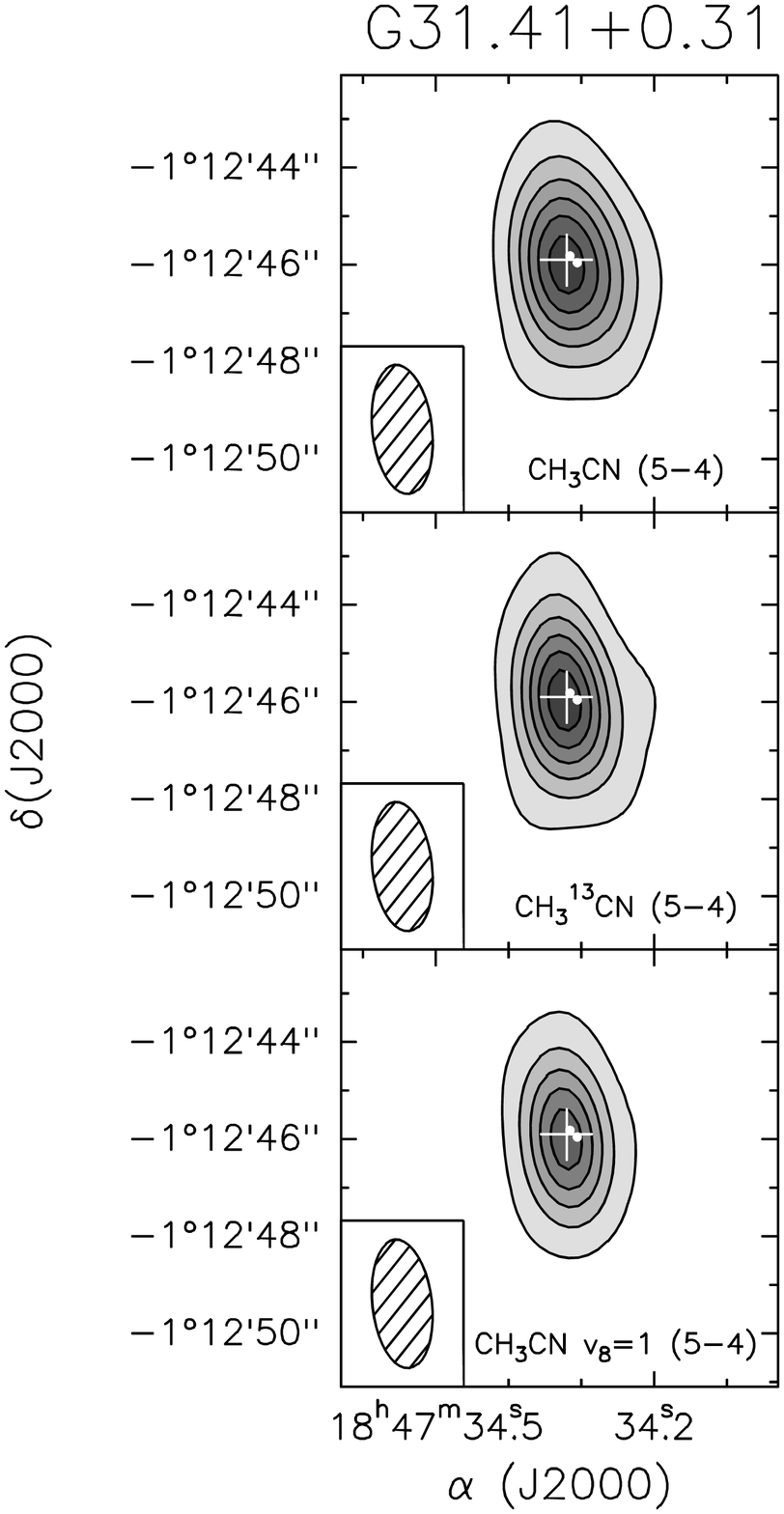}}
\caption{PdBI intensity maps of the \MCN\ (\jcc) ({\it top panel}) and the \MCNII\ ({\it middle panel}) emission averaged under the $K=0$ to 4 components, and of the $v_8=1$ emission ({\it bottom panel}) averaged under the (1,1), (2,1), (0,1), (3,1), (1,$-1$), (4,1), (2,$-1$), (3,$-1$), (4,$-1$), and (1,1) transitions toward G31.41+0.31. The contour levels range 
from 0.1 to 0.70\jy\ in steps of 0.12\jy\ ({\it top panel}), from 0.02 to 0.22\jy\ ({\it middle panel}), and to 0.18\jy\ ({\it bottom panel}) in steps of 0.04\jy. The synthesized beam is shown in the lower left-hand corner.
The white cross marks the position of the 1.4~mm continuum emission peak, and the white dots the 7~mm continuum emission peaks detected by Hofner (private communication).}
\label{g31_ch3cn_3mm}
\end{figure}

Figure~\ref{g31-spectra} shows the spectra of the \MCN\ and \MCNII\ (\jdo) and
(\jcc) lines toward G31.41+0.31. The first vibrational state above the ground
of the \MCN\ (\jcc) line, which is denoted $v_8=1$, is also shown in this
figure. The spectra have been taken integrating the emission over the $3\sigma$
contour level area. Figure~\ref{g31_ch3cn_1mm} shows the maps of the \MCN\
(\jdo) emission averaged under the $K=1$ and 2 components and under the $K=8$
component, and the \MCNII\ (\jdo) emission averaged under the $K=1$ and 2
components. Figure~\ref{g31_ch3cn_3mm} shows the maps of the \MCN\ (\jcc) and
the \MCNII\ (\jcc) emission averaged under the $K=0$ to 4 components, and the
\MCN\ (\jcc) $v_8=1$ emission averaged under all observed transitions.

Several $K$-components of the different rotational transitions of methyl
cyanide are clearly detected toward G31.41+0.31 as can be seen in Fig.~\ref{g31-spectra}. We fitted multiple
$K$-components of a given rotational transition by fixing their separation in
frequency to the laboratory values (see Lovas~\cite{lovas92} and references therein) and forcing their line width to be
identical. As different $K$-components of a same rotational transition fall in
different PdBI correlator units with different spectral resolutions, sometimes
we did not fit simultaneously all of the lines for a given rotational
transition, but  groups of them. The groups were of 5 lines at maximum because
CLASS, which is the software package used to analyze the lines, only allows to
fit 5 lines simultaneously. In addition, in a few cases, the line width in the
fit was fixed to the value obtained for other $K$-components of the same
transition; otherwise, due to the low spectral resolution with which those
transitions have been observed, the line widths obtained when fitting the
profiles were unrealistically wide. The fact of fixing the line width for groups of lines with different $K$-components fitted simultaneously does not affect  significantly the results and conclusions derived from the analysis of the lines. The profiles show no evidence of self-absorption, are quite
symmetric and can be reasonably fitted by a Gaussian profile. The line
parameters of  the ground state transitions of  \MCN\ and \MCNII\ are given in
Table~\ref{table_ch3cn_g31}, while those of the $v_8=1$ are given in
Table~\ref{table_v8_g31}. Note that in some cases the lines are so blended with
other methyl cyanide components (e.g.\ the \MCNII\ $K=0$ with \MCN\  $K=5$ for
the (\jdo) transition), or with other molecular lines such as $^{13}$CO (\jdu),
that it has been impossible to derive their parameters.  As can be seen in Tables~\ref{table_ch3cn_g31} and
\ref{table_v8_g31} there is a good agreement between the systemic velocities
derived from all transitions (\Vlsr~$\simeq 97~$\kms). Regarding the
line widths, we found a similar value,  FWHM~$\simeq 8$~\kms, for \MCN\ (\jdo),
\MCNII\ (\jdo), and \MCN\ (\jcc) $v_8=1$. On the other hand, for \MCN\ (\jcc)
and \MCNII\ (\jcc) we found values of $\sim 10$ and $\sim 6$~\kms, 
respectively.
Assuming that the line profiles are Gaussian, the ratio of the observed line width (broadened by the line optical depth) and the intrinsic one, can be written as  
\begin{equation}
\frac{\Delta V_{\rm obs}}{\Delta V_{\rm int}} = \frac{1}{\sqrt{\ln 2}} \sqrt{-\ln\bigg[-\frac{1}{\tau_0}\,\ln\bigg(\frac{1+{\rm e}^{-\tau_0}}{2}\bigg)\bigg]}, 
\label{equ1}
\end{equation}
where $\tau_0$ is the optical depth at the center of the line. In practice, Eq.~(\ref{equ1}) gives the line width ratio between an optically
thick and an optically thin transition of the same species, where
$\tau_0$ corresponds to the optical depth of the thick line.
One may hence apply Eq.~(\ref{equ1}) to the $K$ lines of \MCN\ (\jcc) and  \MCNII\ (\jcc). The optical depth $\tau_0$ can be derived from the brightness temperature ratio between lines of the
two species with the same $K$ (assuming a relative abundance [\MCN]/[\MCNII]=50; see Wilson \& Rood \cite{wilson94}): using this in Eq.~(\ref{equ1}) one obtains the line width ratio between \MCN\ and \MCNII\ expected if line broadening is due only to optical depth effects. Such a ratio turns out to range between 1.8
and 2.1, roughly consistent with that measured directly from the line
profiles: $1.6\pm0.02$. We conclude that the \MCN\ lines are wider than the
\MCNII\ ones as a consequence of the larger optical depth, although one cannot exclude that different
methyl cyanide transitions are tracing gas with slightly different physical
conditions as well.

\begin{table*}
\caption[] {\MCN\ and \MCNII\  line parameters for G31.41+0.31}
\label{table_ch3cn_g31}
\begin{tabular}{lccccc}
\hline
\multicolumn{1}{c}{Line} &
\multicolumn{1}{c}{$V_{\rm LSR}$} &
\multicolumn{1}{c}{FWHM} &
\multicolumn{1}{c}{$T_{\rm B}^{\rm a}$} &
\multicolumn{1}{c}{$\int{T_{\rm B}\,{\rm d}V}$} &
\multicolumn{1}{c}{rms} \\
\multicolumn{1}{c}{$K$} &
\multicolumn{1}{c}{(\kms)} &
\multicolumn{1}{c}{(\kms)} &
\multicolumn{1}{c}{(K)} & 
\multicolumn{1}{c}{(K \kms)} &
\multicolumn{1}{c}{(K \kms)} \\
\hline
&&&\MCN\ (\jcc)  \\
\hline
0 &$96.29\pm0.03$ &$9.77\pm0.04$ &$19.0\pm1.3$ &$198\pm3$ &0.8\\
1 &" &" &$15.9\pm1.3$ &$165\pm3$ &" \\
2 &" &" &$21.6\pm1.3$ &$225\pm2$ &" \\
3 &" &" &$22.2\pm1.3$ &$231\pm2$ &" \\
4 &" &" &$14.2\pm1.3$ &$148\pm2$ &" \\
\hline
&&&\MCNII\ (\jcc)  \\
\hline
0 &$97.41\pm0.03$ &$6.11\pm0.05$ &$5.83\pm0.46$ &$37.9\pm0.6$ &0.24\\
1 &" &" &$6.45\pm0.46$ &$41.9\pm0.6$ &" \\
2 &" &" &$4.76\pm0.46$ &$31.0\pm0.6$ &" \\
3 &" &" &$5.41\pm0.46$ &$35.2\pm0.6$ &" \\
4 &$97.41^{\rm b}$ &$6.11^{\rm b}$ &~$2.17\pm0.18^{\rm c}$ &~$14.1\pm1.2^{\rm d}$ &0.08\\
\hline
&&&\MCN\ (\jdo)  \\
\hline
0 &$96.24\pm0.06$ &$8.07\pm0.08$ &$36.7\pm1.4$ &$315\pm8$ &2.9\\
1 &" &" &$31.3\pm1.4$ &$269\pm8$ &"\\
2 &" &" &$39.0\pm1.4$ &$335\pm8$ &" \\
3 &" &" &$40.4\pm1.4$ &$347\pm8$ &"\\
4 &$96.84\pm0.09$ &$8.10\pm0.12$ &$34.9\pm1.2$ &$301\pm7$ &2.5\\
5 &" &" &$26.9\pm1.2$ &~$232\pm11$ &" \\
6 &$96.78\pm0.16$ &$8.00^{\rm b}$ &$24.3\pm3.0$ &$207\pm7$ &1.5\\
7 &$97.15\pm0.35$ &$8.00^{\rm b}$  &$16.1\pm3.0$ &~$137\pm16$ &3.6\\
8 &" &" &$10.9\pm3.0$ &~~$93\pm18$ &" \\
9 &$97.85\pm1.28$ &$8.00^{\rm b}$ &(e) &(e) &3.4\\
10 &" &" &~$3.4\pm0.7$ &~$29\pm22$ &"\\
11 &$97.12\pm0.65$ &$8.00^{\rm b}$ &~$1.0\pm1.6$ &~~$8\pm17$  &2.6 \\
\hline
&&&\MCNII\ (\jdo)  \\
\hline
0 &$96.84\pm0.09$ &$8.10\pm0.12$ &(f) &(f) &2.5\\
1 &" &" &$14.2\pm2.5$ &$122\pm8$ &"\\
2 &" &" &$12.9\pm2.5$ &$111\pm6$ &" \\
3 &$96.78\pm0.16$ &$8.00^{\rm b}$ &$19.8\pm3.0$ &$169\pm7$ &1.5 \\
4 &" &" &$9.0\pm3.0$ &$76.3\pm6.6$ &" \\
5 &" &" &$5.0\pm3.0^{\rm g}$ &$42.2\pm7.2^{\rm g}$ &" \\
6 &" &" &$3.8\pm3.0$ &$32.1\pm6.6$ &" \\
7 &$97.85\pm1.28$ &$8.00^{\rm b}$ &$2.6\pm0.7$ &$22\pm21$ &3.4\\
8 &$97.12\pm0.65$ &$8.00^{\rm b}$ &$15.0\pm1.6^{{\rm h}}$ &$127\pm17^{\rm h
}$ &2.6\\
9 &" &" &$6.4\pm1.6$  &$55\pm17$ &" \\
\hline

\end{tabular}
   
  (a) Brightness temperature integrated over a $3\sigma$ emission level area. \\
  (b) This parameter was held fixed in the fit. \\
  (c) Computed from the integrated line intensity. \\  
  (d) Measured with the command print area of CLASS. \\
  (e) Blended with $^{13}$CO (\jdu). \\
  (f) Blended with \MCN\ $K=5$. \\
  (g) Blended with \MCN\ $K=7$. \\
  (h) Dominated by wing emission of  $^{13}$CO (\jdu).   
  
\end{table*}

\begin{table*}
\caption[] {\MCN\ (\jcc) $v_8=1$  line parameters for G31.41+0.31}
\label{table_v8_g31}
\begin{tabular}{lccccc}
\hline
\multicolumn{1}{c}{Line} &
\multicolumn{1}{c}{$V_{\rm LSR}$} &
\multicolumn{1}{c}{FWHM} &
\multicolumn{1}{c}{$T_{\rm B}^{\rm a}$} &
\multicolumn{1}{c}{$\int{T_{\rm B}\,{\rm d}V}$} &
\multicolumn{1}{c}{rms} \\
&\multicolumn{1}{c}{(\kms)} &
\multicolumn{1}{c}{(\kms)} &
\multicolumn{1}{c}{(K)} & 
\multicolumn{1}{c}{(K \kms)} &
\multicolumn{1}{c}{(K \kms)} \\
\hline
(1,1) &$97.45\pm0.41$ &$8.02\pm0.44$ &$5.15\pm0.24$ &$44\pm5$ &1.0\\
(2,1) &" &" &$4.69\pm0.24$ &$40\pm5$ &" \\
(0,1) &" &" &$4.62\pm0.24$ &$39\pm5$ &" \\
(3,1) &" &" &$3.65\pm0.24$ &$31\pm5$ &" \\
$(1,-1)$ &" &" &$4.12\pm0.24$ &$35\pm5$ &" \\
$(4,1)$ &$97.45\pm0.50$ &$8.71\pm0.69$ &$3.17\pm0.31$ &$29\pm7$ &1.2\\
$(2,-1)$ &" &" &$4.95\pm0.31$ &$46\pm7$ &" \\
$(3,-1)$ &" &" &$3.27\pm0.31$ &$30\pm7$ &"\\
$(4,-1)$ &" &" &$1.63\pm0.31$ &$15\pm6$ &"\\
$(1,1)$ &$97.45^{\rm b}$ &$8.71^{\rm b}$ &$4.72\pm0.12^{\rm c}$ &~$39\pm1^{\rm d}$ &0.09\\
\hline

\end{tabular}

 (a) Brightness temperature integrated over a $3\sigma$ emission level area. \\
 (b) This parameter was held fixed in the fit. \\
 (c) Computed from the integrated line intensity. \\  
 (d) Measured with the command print area of CLASS. \\  
\end{table*}

\begin{figure}
\centerline{\includegraphics[angle=-90,width=7cm]{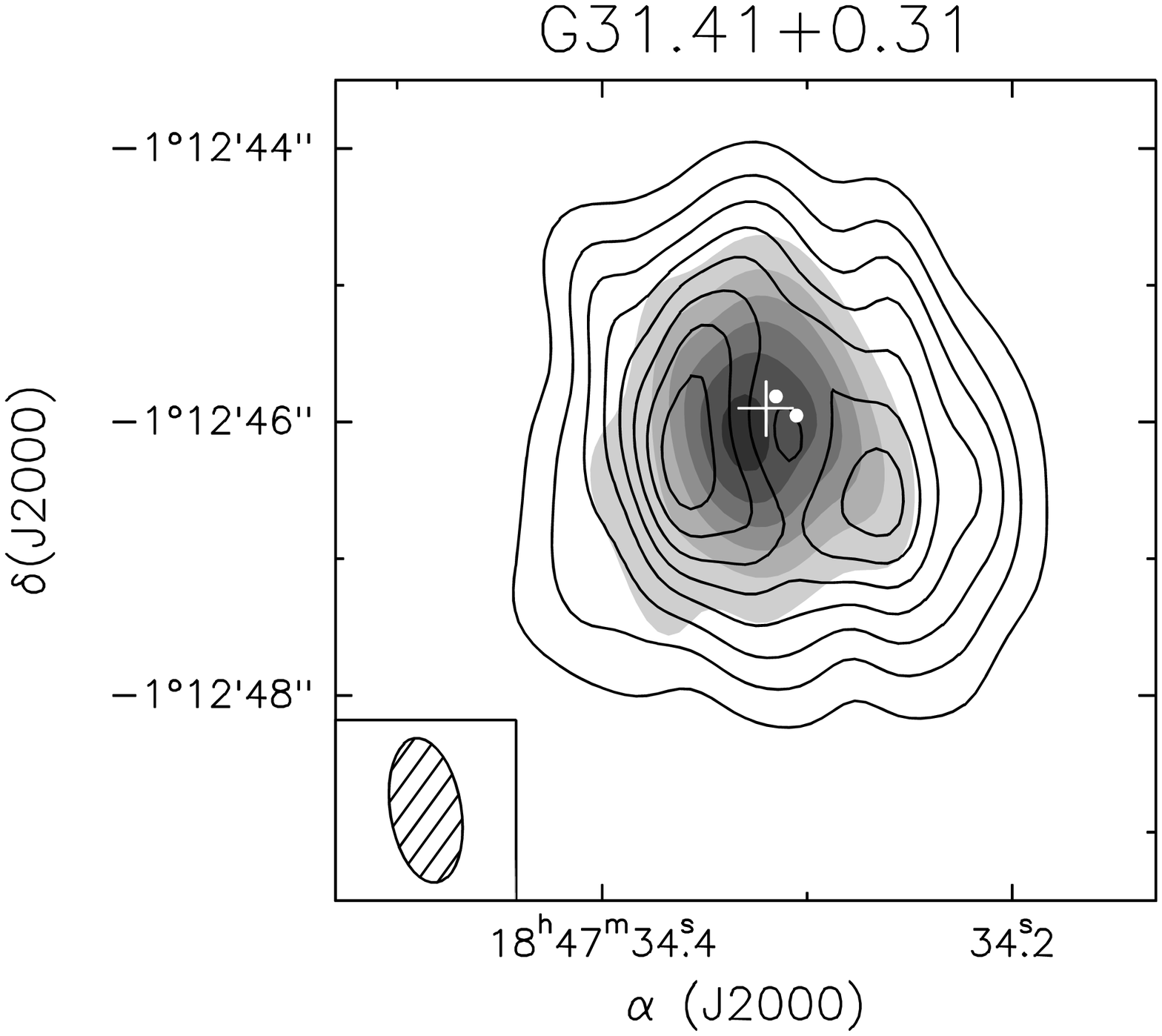}}
\vspace{0.5cm}
\caption{Overlay of the PdBI \MCN\ (\jdo) emission averaged under the $K=0$, 1, and 2 components ({\it contours}) on the emission under $K=8$ ({\it greyscale}) toward G31.41+0.31. Contour levels
 range from 0.1 to 0.94~Jy\,beam$^{-1}$ in steps of 0.12~Jy\,beam$^{-1}$. Greyscale levels range from 0.1 to 0.70~Jy\,beam$^{-1}$ in steps of 0.12~Jy\,beam$^{-1}$\,\kms. The synthesized beam is shown in the lower left-hand corner. The white cross marks the position of the 1.4~mm continuum emission peak, and the white dots the 7~mm continuum emission peaks detected by Hofner (private communication).}
\label{g31_k8_k012}
\end{figure}

 As can be seen in Fig.~1 of Paper~I, and in Fig.~\ref{g31_ch3cn_1mm}, the
\MCN\ $K=1$ and 2 components seem to trace a toroidal structure with the
millimeter continuum emission peaking at the central dip. To explain this
morphology two possible scenarios are suggested: $(i)$ a dramatic \MCN\
abundance drop in the central region of the core; or $(ii)$ self-absorption,
due to the high optical depth at the center of the core and the existence of a
temperature gradient. Figures~\ref{g31_ch3cn_1mm} and \ref{g31_k8_k012} show
that the different $K$-components do not trace the same material, and that for
large values of $K$ the \MCN\ emission peaks closer to the center. In fact, as
shown in Fig.~\ref{g31_k8_k012}, the $K=8$ emission fills the central dip
visible in the $K=0$, 1, and 2 emission. The \MCNII\ $K=$1 and 2 emission is
also found closer to the center than the \MCN\ $K=$1 and 2 emission. Thus, the
lack of emission at the center of the core occurs only for the lower
$K$-components of \MCN, unlike the higher $K$-components, \MCNII, and continuum
emission. This indicates that the \MCN\ molecules are not destroyed at the
center of the core.  As can be seen in Fig.~\ref{g31-spectra}, the \MCN\ (\jdo)
$K=0$ to 4 lines show similar intensities, suggestive of high optical depth.
Note that low-$K$ transitions of \MCNII, which have excitation energies very
similar to those of the corresponding \MCN\ $K$ lines, trace a slightly inner
region: this is an opacity effect as the less abundant isotopomer traces higher
column densities. In order to confirm a possible opacity effect, we computed
the optical depth at each point using the $K=2$ components of \MCN\ and \MCNII,
assuming that \MCN\ is optically thick and that the excitation temperature is
the same for both transitions. A map of the optical depth of the \MCN\ main
species, $\tau_{12}$, shows that the opacity increases toward the center of the
core: it ranges from 10 to 50 for $K=2$, and from 5 to 20 for $K=6$. In
conclusion, the most plausible explanation for the apparent toroidal geometry
of the \MCN\ map in Fig.~\ref{g31_ch3cn_1mm} is self-absorption, which is
caused by the high optical depth and the temperature gradient in the core. In
Sect.~\ref{g31-tempgrad} we discuss in more detail this temperature gradient.  

\subsection{Discussion}

\subsubsection{Temperature gradient}
\label{g31-tempgrad}

\begin{figure}
\centerline{\includegraphics[angle=-90,width=8cm]{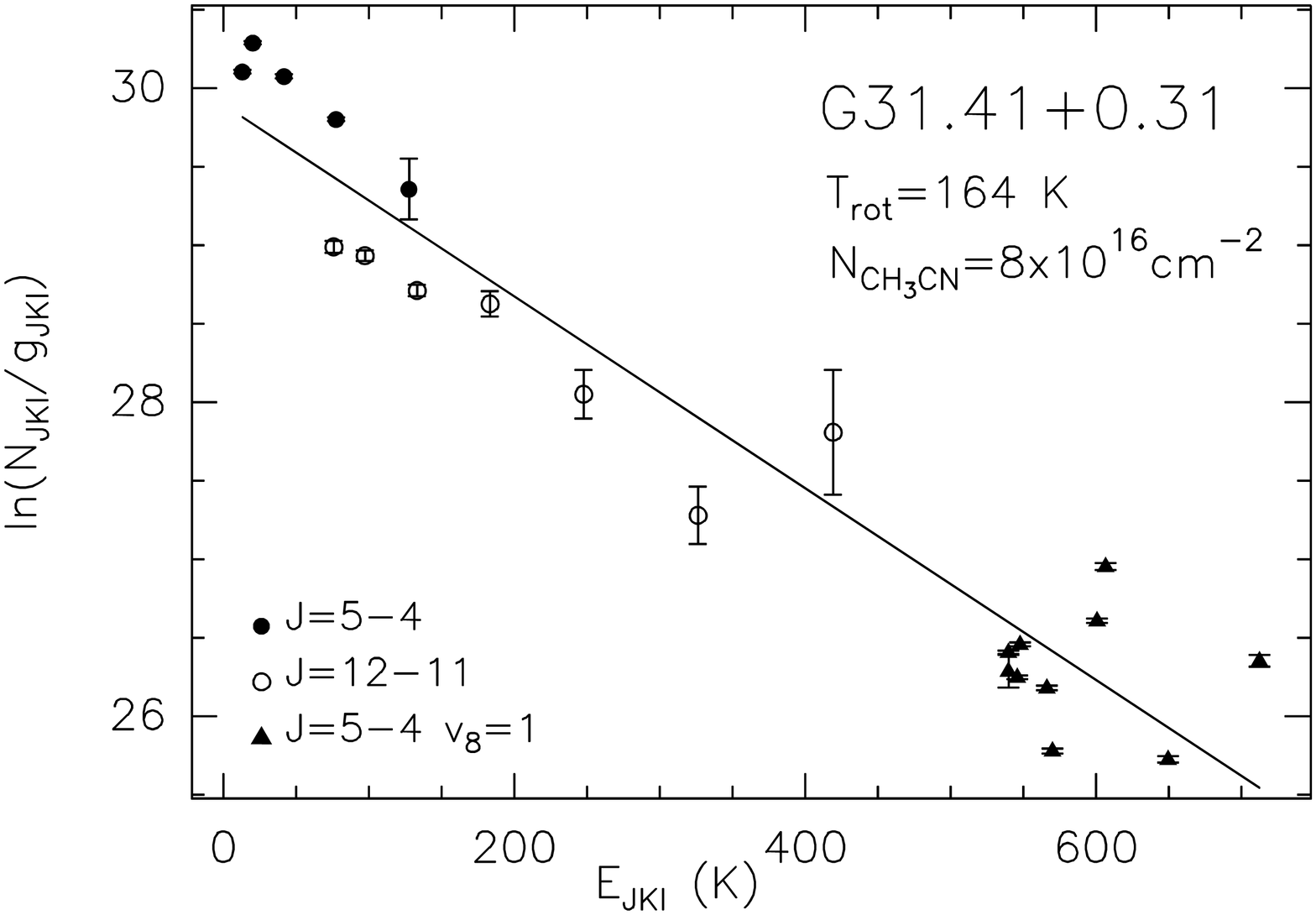}}
\vspace{0.3cm}
\caption{Rotation diagram for G31.41+0.31 with superimposed fit. Filled circles, open circles, and  filled triangles correspond to the \MCNII\ (\jcc), the \MCNII\ (\jdo), and the \MCN\ (\jcc) $v_8=1$ transition, respectively.}
\label{g31-boltz}
\end{figure}

In Sect.~\ref{g31-ch3cn} we have illustrated the existence of a temperature gradient in the G31.41+0.31 core. Here we want to estimate the temperature and its variation across the core. Following Olmi et al.~(\cite{olmi93}) we derived estimates for the rotational temperature,
$T_{\rm rot}$, and the total methyl cyanide column density, $N_{\rm CH_3CN}$,
by means of the rotation diagram (RD) method, which assumes that the molecular
levels are populated according to LTE conditions at a single temperature
$T_{\rm rot}$. In the high density limit where level populations are
thermalised, one expects that $T_{\rm rot}= T_{\rm k}$, the kinetic
temperature. The \MCN\ ground level transitions appear to be optically thick,
as suggested by the ratio between main species and isotopomer. As already mentioned in the previous section, $\tau_{12}$, which is position depending and has been evaluated at the systemic velocity, ranges from 10 to 50 for $K=2$, and from 5 to 20 for $K=6$ in the core. Thus, in the fit performed to the Boltzmann
plot we used only the transitions of \MCNII, and
those from the \MCN\ $v_8=1$ state (see Fig.~\ref{g31-boltz}). All the
spectra were integrated over a region of $\sim 4\farcs7$ in diameter ($\sim 
17~{\rm arcsec}^2$),
in order to include all the emission inside the $3\sigma$ contour level for the (\jcc)
transition. Assuming that also the emission due to the vibrational state is optically thin and a relative
abundance [\MCN]/[\MCNII]=50 (see Wilson \& Rood \cite{wilson94}), the $T_{\rm rot}$ and source averaged $N_{\rm CH_3CN}$ derived by taking
into account all transitions mentioned above are $164\pm5$~K and
$8\times10^{16}$~\cmq, respectively. We also estimated $T_{\rm rot}$ and
$N_{\rm CH_3CN}$ from each set of \mbox{$J+1\rightarrow J$} lines separately. The rotation
temperatures estimates are $135\pm2$ and $220\pm16$~K, and the column densities
$9\times10^{16}$ and $7\times10^{16}$~\cmq\, for the (\jcc) and (\jdo)
transitions, respectively. These values of $T_{\rm rot}$ are consistent
with those
derived from \MCN\ and \MCNII\ (\jsc) by Olmi et al.~(\cite{olmi96a})  for the
compact component of the molecular clump ($\sim$ 1--2$\arcsec$), called ``core", and the most extended one ($\sim 10$--20$\arcsec$), called ``halo", which are  230 and 130~K, respectively. The fact that the $T_{\rm rot}$ that we derived by using only the (\jdo)
transitions is close to the value derived by Olmi et al.~(\cite{olmi96a}) for
the ``core", and that derived using only the (\jcc) lines is closer to the value for the
``halo", could be indicating that the highest excitation transitions are tracing
material closer to the embedded source, while the lowest  excitation transitions
are tracing material in the outer shell of the core. From this analysis, it is clear that the
temperatures in G31.41+0.31 are very high, confirming that the methyl cyanide
emission comes from hot molecular gas. Regarding the column densities, the
estimates of the order of $\sim 10^{17}$~\cmq\ are consistent with those
obtained by Olmi et al.~(\cite{olmi96a}, \cite{olmi96b}).

\begin{figure}
\centerline{\includegraphics[angle=0,width=8.5cm]{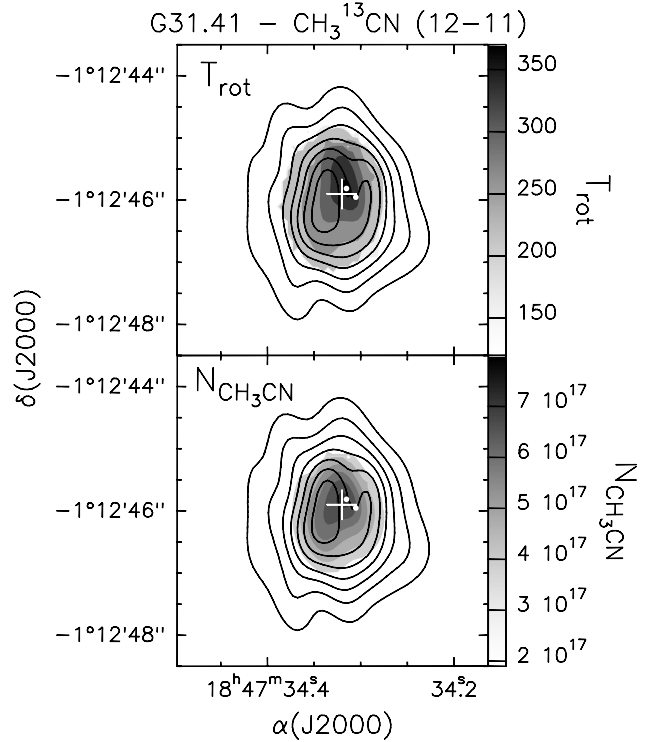}}
\caption{Overlay of the PdBI \MCNII\ (\jdo) emission averaged under the $K=1$ and 2 components ({\it contours}) on the $T_{\rm rot}$ map ({\it top panel}) and the $N_{\rm CH_3CN}$ map ({\it bottom panel}) in greyscale, derived by fitting  the Boltzmann plots at each position, toward G31.41+0.31. The contour levels range from 0.1 to 0.82\jy\ in steps of 0.12\jy. Greyscale levels range from 100 to 350~K by 50~K ({\it top panel}), and from 2 to 
$7\times 10^{17}$~\cmq\ by $1\times 10^{17}$~\cmq\ ({\it bottom panel}). The white cross marks the position of the 1.4~mm continuum emission peak, and the white dots the 7~mm continuum emission peaks detected by Hofner (private communication).}
\label{g31_trot_ntot}
\end{figure}

In order to establish the existence of a temperature gradient in the core, as
suggested by Figs.~\ref{g31_ch3cn_1mm} and \ref{g31_k8_k012} (see
Sect.~\ref{g31-ch3cn}), we computed a map of $T_{\rm rot}$ by fitting the
Boltzman plots at each position (pixel) in the core. The fits were performed
with the RD method to the \MCNII\ $K$=1, 2, 3, 4, 5, and 6 components ($K$=0
was not used because too blended with \MCN\ $K$=5). Figure~\ref{g31_trot_ntot}
shows the resulting map of $T_{\rm rot}$, as well as the total column density
map, $N_{\rm CH_3CN}$, derived with the same method; for the sake of comparison
the \MCNII\ (\jdo) emission averaged under the $K=1$ and 2 components is also
shown. This figure clearly confirms that the temperatures reached toward the 
center of the core, where the millimeter continuum source is embedded and the
lower $K$-components methyl cyanide emission shows a dip, are very high ($\sim
350$~K), while the temperatures toward the peak of \MCNII\ are lower
($\sim200$--250~K). The temperature at the position of the peaks of the torus
seen in the \MCN\ emission (see Fig.~\ref{g31_k8_k012}) is even lower, $T_{\rm
rot}\sim 150$~K. For this source Cesaroni et al.~(\cite{cesa98})  have derived
a temperature gradient of the type $T \propto R^{-3/4}$ using NH$_3$(4,4): 
this is consistent with our findings $350\sim 150\times3^{3/4}$, where 3 is
roughly the value of the distance where the temperature is 150~K divided by the
distance where the temperature is 350~K. The $N_{\rm CH_3CN}$ map also shows
that the gas column density increases toward the center, with the peak at the
position of the millimeter continuum source, confirming that the \MCN\ gas
is present also in the central region of the core. 

The geometry of the G31.41+0.31 core seems to be the result of the combination
of two effects: the existence of a temperature gradient in the core plus opacity. In Sect.~\ref{g31-ch3cn} we checked for a possible opacity effect by
computing the optical depth at each point assuming, as a first approximation, a
single excitation temperature. However, because of the existence of a
temperature gradient in the core, the optical depth can be better measured
taking it into account. In order to better fit the data and understand the
geometry of the core we computed models which solve the radiative transfer
equation. These models are discussed in the next section.

\subsubsection{Velocity field: the rotating toroid}
\label{g31-models}

In Paper~I we detected a clear velocity gradient in the G31.41+0.31 core
by simultaneously fitting multiple \MCN\ (\jdo) $K$-components at each position
where \MCN\ is detected. This velocity gradient is perpendicular to the
molecular outflow detected in the region by Olmi et al.~(\cite{olmi96a}), and
as already discussed in Paper~I, the most plausible explanation for  such a
velocity gradient is that the core is undergoing rotation about the outflow
axis. Also Gibb et al.~(\cite{gibb04}) have studied the velocity field  in
G31.41+0.31 by  analyzing the C$^{18}$O and H$_2$S line emission. By mapping
the H$_2$S line wings, these authors have detected a velocity gradient in the
same direction as our \MCN\ gradient, but have interpreted this as a bipolar
outflow, with the blueshifted emission toward the west and the redshifted
toward the east. To our knowledge a high-density tracer such as \MCN\ has never
been found to trace outflows: instead there are clear examples of \MCN\
emission tracing rotating disks, for example IRAS 20126+4104 (Cesaroni et
al.~\cite{cesa99}) or G24.78+0.08 (Paper I); in the latter the velocity
gradient seen in \MCN\ in core A1 is clearly perpendicular to the bipolar
outflow mapped in $^{12}$CO (Furuya et al.\cite{furuya02}). We hence believe that the interpretation given in Paper~I for the \MCN\ emission in G31.41+0.31 is to be preferred to that of Gibb et al.~(\cite{gibb04}), and consider the velocity gradient detected a signature of rotation. Only high-angular resolution observations in unambiguous outflow tracers such as $^{12}$CO may definitely prove which interpretation is correct.

The velocity shift measured over an extent of $\sim 17000$~AU is $4.2$~\kms\
(Paper~I), which corresponds to a velocity gradient of about 50~\kms\
pc$^{-1}$. The size and the mass of $\sim 500~M_\odot$ suggest that this toroid could be hosting not a single massive YSO but a cluster of YSOs. This is also suggested by the continuum emission at 7~mm toward the
center, which reveals two point-like sources oriented in the  NE--SW direction
(Hofner, private communication), i.e.\ just along the toroid plane. It is hence
tempting to speculate that collapse in the toroid has led to their formation.

\begin{figure}
\centerline{\includegraphics[angle=0,width=8.5cm]{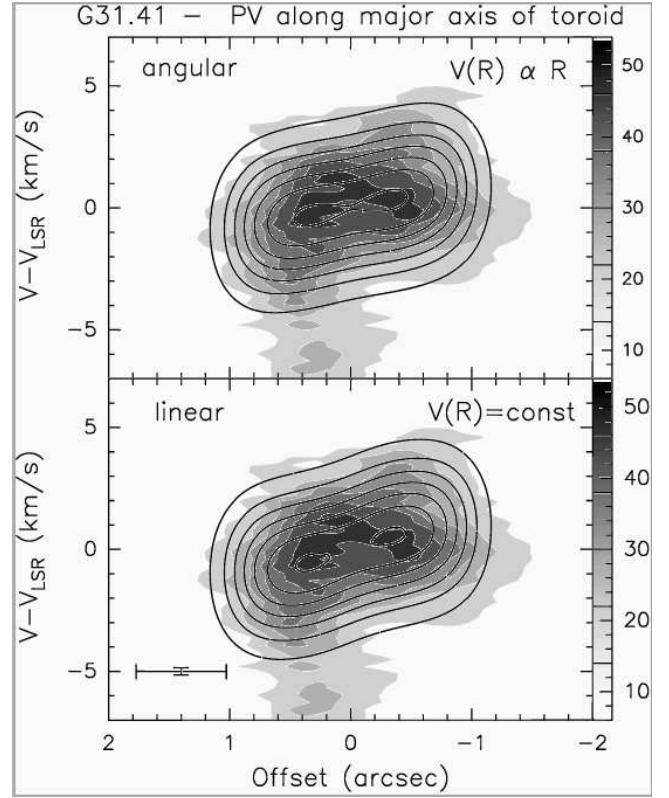}}
\caption{Overlay of the PV plot of the \MCN\ (\jdo) $K=3$ emission ({\it greyscale}) along the major axis of the toroid (P.A. $\sim 70\degr$) in G31.41+0.31, and the synthetic emission PV plot of a model ({\it contours}) with constant angular velocity ({\it top panel}), and with constant rotation velocity ({\it bottom panel}). The {\it y}-axis plots the difference between the measured velocity and the systemic velocity, which is 97~\kms\ for G31.41+0.31. The contour levels range from 6 to 54~\jy\ in steps of 8~\jy. The error bars in the lower left-hand corner indicate the angular and spectral resolution of the data and the models.}
\label{g31_pv_models}
\end{figure}

The kinematics of the gas toward the core of G31.41+0.31 can be seen in the
position-velocity (PV) cut of the \MCN\ (\jdo) $K=3$ emission along a direction
(P.A. $\sim 70\degr$) where the velocity gradient is maximum
(Fig.~\ref{g31_pv_models}). Clearly, Keplerian rotation, which has been
detected toward lower luminous objects such as IRAS~20126+4104 (Cesaroni et
al.~\cite{cesa97}, \cite{cesa99}; Zhang et al.~\cite{zhang98}), NGC~7538S
(Sandell et al.~\cite{sandell03}), or M17 (Chini et al.~\cite{chini04}), is not
possible as the mass of the toroid is much larger than any reasonable stellar
mass. In fact, the PV cut shows no hint of Keplerian rotation. However, it is
interesting to see that according to theoretical models (Galli, private
communication), discriminating between a velocity field with constant rotation
velocity or with constant angular velocity should help to assess whether
magnetic braking, and thus magnetic field, plays a crucial role during the
process of disk, and hence star, formation. In the absence of magnetic field, a
self-gravitating isothermal quasistatic equilibrium situation is reached when
the rotation velocity is constant, $v =$ constant ($\gamma = 0$). In case that
magnetic field is present, it couples the core with the surrounding medium  by
means of magnetic braking. This coupling yields a constant angular velocity, $v
\propto R$ ($\gamma = 1$), in the whole core, which rotates as a rigid body and
leads by ambipolar diffusion to an equilibrium situation. Thus, in order to
discriminate between the two possible scenarios, which could give an indication
of the stability of the cores and a hint on their evolutionary stage, we
modeled the \MCN\ emission by assuming that the emission arises from a
disk-like structure seen edge-on by the observer with an internal radial
velocity field ($v\propto R^\gamma$); a constant rotation velocity field and a
constant angular velocity field.

The models were computed by adopting a power law dependence on the distance
from the core center also for the temperature $T\propto R^{-q}$ of the emitting
gas. According to accretion disk theory for geometrically thin disks the
dependence of the temperature on the radius is $T\propto R^{-3/4}$ (see
Natta~\cite{natta00} and references therein). Although geometrically thick, these toroids resemble more 2-D structures like circumstellar disks than 3-D spherically symmetric cores. Therefore, we have adopted a temperature profile $T\propto R^{-3/4}$ in our models. This is also indicated by the findings of Cesaroni et al.~(\cite{cesa98}), who have properly modeled the NH$_3$(4,4) emission toward G31.41+0.31 with  $T \propto R^{-3/4}$. Regarding the density distribution, for
simplicity we adopted constant density in the core. The synthetic PV diagrams
were computed along the projected major axis of the toroid, and convolved with
a Gaussian in the PV plane to obtain the same angular and spectral resolutions
as those of the data. The parameters of the models are the inner and outer
radius of the toroid, $R_{\rm inn}$ and $R_{\rm out}$, respectively, a
parameter related to the column density at the peak of the emission, the
observed line width, $\Delta V$, the power-law index $p$ of the density, the
temperature at the outer radius, $T_{\rm out}$, the power-law index $q$ of the
temperature, and the rotation velocity measured at the edge of the toroid,
$v_{\rm rot}$. The initial guesses for the input parameters such as the size of
the toroid or the rotation velocity were derived from the observations (see
Paper~I).

\begin{figure}
\centerline{\includegraphics[angle=-90,width=7.5cm]{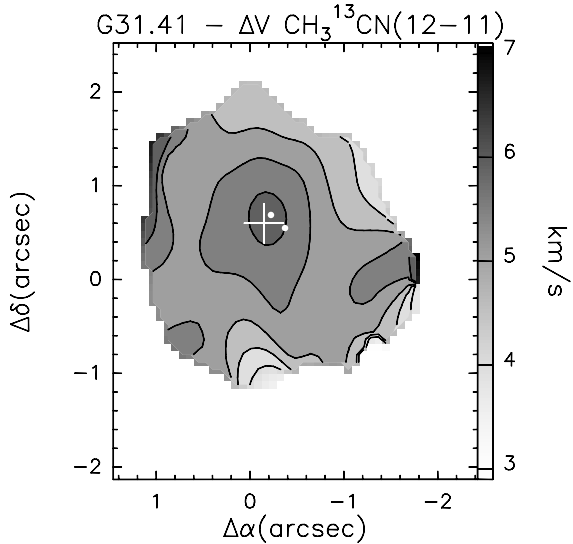}}
\caption{Line width plot of \MCNII\ (\jdo) toward G31.41+0.31. The contour levels range from 3 to 7~\kms\ in steps of 1~\kms. The white cross marks the position of the 1.4~mm continuum emission peak, and the white dots the 7~mm continuum emission peaks detected by Hofner (private communication).}
\label{g31-linewidth}
\end{figure}

\begin{table*}
\caption[] {Best fit model parameters for G31.41+0.31 and core A1 in G24.78+0.08}
\label{table_models}
\begin{tabular}{lcccccccccc}
\hline
&\multicolumn{2}{c}{$R_{\rm inn}$} &&
\multicolumn{2}{c}{$R_{\rm out}$} &
\multicolumn{1}{c}{$T_{\rm out}$} &
\multicolumn{1}{c}{$N_{\rm CH_3CN}^{\rm peak~~a}$} &
\multicolumn{1}{c}{$v_{\rm rot}$} &
\multicolumn{1}{c}{$\Delta V$} &
\multicolumn{1}{c}{$q$} \\
\cline{2-3}
\cline{5-6}
\multicolumn{1}{c}{Core} &
\multicolumn{1}{c}{($''$)}&
\multicolumn{1}{c}{(AU)} &&
\multicolumn{1}{c}{($''$)} & 
\multicolumn{1}{c}{(AU)} &
\multicolumn{1}{c}{(K)} & 
\multicolumn{1}{c}{(cm$^{-2}$)} &
\multicolumn{1}{c}{(\kms)} &
\multicolumn{1}{c}{(\kms)} \\
\hline
G31 &0.17 &1340 &&1.7 &13400 &100 &$5\times10^{15}$ &1.7 &6.8  &$-0.75$ \\
G24  A1 &0.30 &2300 &&1.0 &7700 &100 &$5\times10^{15}$ &2.0 &5.6 &$-0.75$\\ 
\hline

\end{tabular}

 (a) Column density at the peak of the \MCN\ emission.

\end{table*}

The observed and synthetic PV diagrams for both scenarios, constant angular
velocity (angular) and constant rotation velocity (linear), can be seen in
Fig.~\ref{g31_pv_models}. From the modeling itself it is not possible to
distinguish between constant angular ($v \propto R$; top panel) or constant rotation velocity ($v =$ constant; bottom panel), because
both velocity fields suitably fit the data. The models were computed with 
and without
temperature gradients. However, only those models with $T\propto R^{-3/4}$ can
fit properly the observed PV plots and reproduce the ``inner hole'' seen in the
data.  A similar result could be also obtained with a bigger $R_{\rm inn}$.
However, in this case the hole at the center of the toroid would be also
visible in the millimeter continuum emission maps. The best fit model parameters are given in
Table~\ref{table_models}. The value of $R_{\rm inn}$ is $0\farcs$17 (1340~AU), 
which is roughly the separation between the two sources detected at 7~mm
(Hofner, private communication). The values of $v_{\rm rot}$ and $R_{\rm out}$
are consistent with those derived directly from the velocity gradient and the
1.4~mm continuum emission at 50\% of the peak, respectively (see Table~1 in
Paper~I). The value of the column density at the peak of the \MCN\ emission,
$N_{\rm CH_3CN}^{\rm peak}$, is $\sim 20$ times lower than the column density
derived by means of the RD method (see previous section). This is due to the fact that the models do not take into account the
clumpiness of the region (i.e.\ a beam filling factor effect). The models
also support the scenario according to which the lack of emission at the center of the toroid is not a geometric
effect but is due to self-absorption, which is caused by the high temperatures and optical depths reached toward the center of the core,
as suggested by the \MCN\ and \MCNII\ integrated emission
plots and by the column density distribution obtained from 
the Boltzman plots  (see Sect.~\ref{g31-ch3cn} and
\ref{g31-tempgrad}). In fact, optically
thin models with a temperature gradient cannot fit the data because in such a case, the external layers of the toroid become too optically thin to absorb
enough photons from the center to reproduce the lack of emission or
self-absorption visible in \MCN. Thus, the models indicate that the emission in
G31.41+0.31 has to be optically thick and that a temperature gradient
is present in the core.

The dynamical mass needed for equilibrium, $M_{\rm dyn}$, has been
derived from the expression $M_{\rm dyn}=\mbox{$v^2_{\rm rot}$}\,R_{\rm out}/
G\, \sin^2i$, where $i$ is the inclination angle assumed to be $90\degr$ for an
edge-on toroid. Note that this expression in Paper~I is wrong: $\sin^2i$ should
be dividing, but the values of $M_{\rm dyn}$ given in Table~1 were correctly
computed. The value obtained is $M_{\rm dyn} \sim 44~M_{\odot}$. The fact that
$M_{\rm dyn}$ is much smaller than the mass of the core suggests, as already
pointed out in Paper~I, that such structure is unstable against gravitational
collapse. Note that the magnetic field required to stabilize it would
be of a few 20--40~mG, values too large to be plausible even in regions as
dense as 10$^8$~cm$^{-3}$ (see Fig.~1 of Crutcher~\cite{crutcher99}). 

\begin{figure*}
\centerline{\includegraphics[width=12cm]{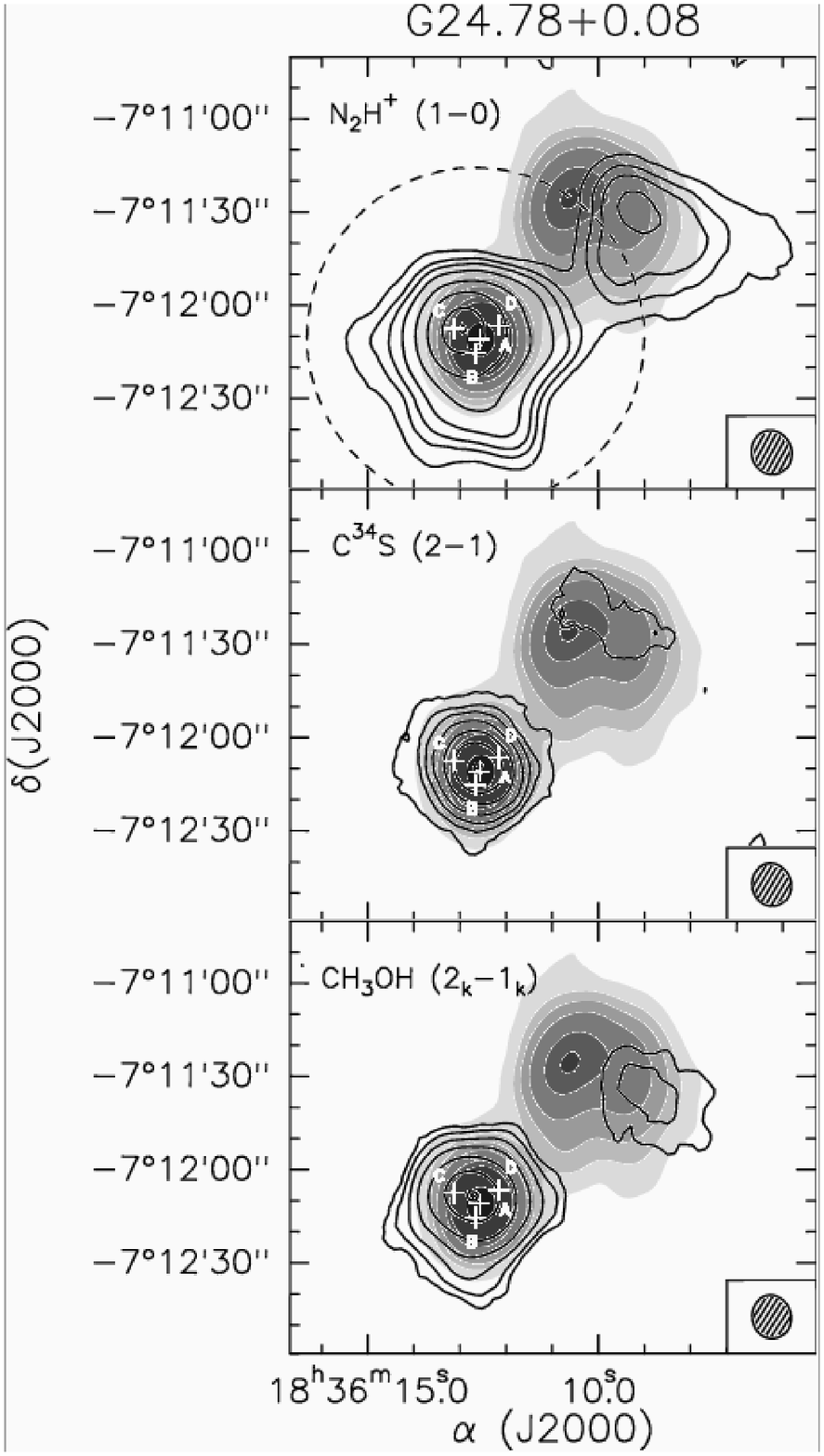}}
\caption{Overlay of the BIMA integrated intensity maps of the N$_2$H$^+$ (\juz) emission ({\it contours;  top panel}), the C$^{34}$S (\jdu) emission ({\it contours; middle panel}), and the \METH\ (\jduk) emission ({\it contours; bottom panel}) on the BIMA 3.16~mm continuum emission ({\it greyscale}) toward G24.78+0.08. The contours are 3, 9, 15, 25, 45, 75, 120, and  150\jykms\ for N$_2$H$^+$, 1.2, 3.6, 6, 10, 18, 30, and 48 \jykms\ for C$^{34}$S, and 3, 6, 12, 24, 48, 96, and 144 \jykms\ for \METH\ (\jduk), and 3, 9, 15, 25, 45, 60, and 120 times 2.5 \mjy\ for the 3.16~mm continuum emission. The dashed circumference in the {\it top panel} represents the BIMA primary beam (50$\%$ attenuation level). The synthesized beams are shown in the lower right-hand corner.
The crosses pinpoint the positions of the four YSOs identified by Furuya et al.\ (\cite{furuya02}).}
\label{g24_nh_cs_ch3oh}
\end{figure*}

In case that the toroid is undergoing collapse the gas should be more perturbed
toward the center, where the 1.4~mm continuum source is embedded, and the
velocity dispersion should increase inward. Thus, one would expect the line
width, $\Delta V$, to increase toward the center of the toroid. Indeed
Fig.~\ref{g31-linewidth} shows  that $\Delta V$, which has been obtained  by
fitting the optically thin \MCNII\ (\jdo) line profiles at each position with a
Gaussian, increases toward the center of the core, where the YSO(s) is
embedded.  One can obtain an estimate of the infall velocity, $v_{\rm inf}$,
assuming that the only contribution to the variation of $\Delta V$ is infall and that the contribution of turbulence is negligible. In such a case, $v_{\rm inf}$ is given by the difference between $\Delta V$ at the position
of the 1.4~mm continuum source and that at the edge of the toroid, $v_{\rm inf}
\simeq (\Delta V_{\rm inn} - \Delta V_{\rm out})/2$. Interestingly, the value
obtained is $v_{\rm inf} \simeq 2$~\kms, consistent with the $v_{\rm rot}$
obtained from the models. Such agreement between $v_{\rm inf}$ and $v_{\rm
rot}$ has been predicted by some recent theoretical studies (see e.g.\ Allen et
al.~\cite{allen03}). Following the expression given in Paper~I we estimated the
accretion rate and obtained $\dot M_{\rm acc}\sim
3\times10^{-2}~M_\odot$~yr$^{-1}$.  Accretion rates that large have been
estimated for the massive Class~0 object IRAS~23385+6053 (Molinari et al.~\cite{molinari98}), and for the parsec-scale clumps in
which high-mass star formation is observed (Fontani et al.~\cite{fontani02}), and could support the theories that
predict that high-mass stars might form through non-spherical accretion with
large accretion rates.

\section{G24.78+0.08}

G24.78+0.08 is a high-mass star forming region that contains a cluster of
massive YSOs in different evolutionary stages. Furuya et al.~(\cite{furuya02})
firstly detected four molecular cores in the region, identified as A, B, C, and
D in Fig.~(\ref{g24_nh_cs_ch3oh}), with two of them, core A and C, powering a
bipolar molecular outflow. When observed with higher angular resolution, the
continuum and methyl cyanide emission toward core A is resolved into two
separate cores, named A1 and A2 in Paper~I. It should be mentioned that
although one cannot rule out the possibility that the continuum and methyl
cyanide peaks in A2 could be the result of the interaction of the outflow with
the surrounding dense gas, we consider them as two distinct cores in our study (see a more detailed discussion in Sects.~\ref{nature_A} and \ref{models_g24}).

\subsection{Results}

In this section we analyze the line and continuum observations carried out with the BIMA and PdBI interferometers toward G24.78+0.08. Subsequently to the lower angular resolution BIMA data, we present the higher angular resolution PdBI data. By presenting the results in such a  way, the analysis of G24.78+0.08 starts with the study of the large scale molecular clumps, and then continues with the study of the smaller scale embedded cores.

\subsubsection{\nh, \cs\ and \METH}
\label{bima_molecularclumps}

\nh\ (\juz), \cs\ (\jdu), and \METH\ (\mbox{2$_{-1}$--1$_{-1}$}) E, (\mbox{2$_{0}$--1$_{0}$}) A$^+$, (\mbox{2$_{0}$--1$_{0}$}) E  were observed toward the hot core
G24.78+0.08 with the BIMA interferometer. Figure~\ref{g24_nh_cs_ch3oh} shows the
integrated emission around the systemic velocity of 111~\kms, overlaid on the 3.16~mm continuum emission, observed also with BIMA. As may be seen in
this figure, a well defined molecular core is found in association with
G24.78+0.08, although the low angular resolution of the BIMA maps ($\theta \sim
14\arcsec$) makes it impossible to disentangle the emission of each 
core identified in the region by Furuya et al.~(\cite{furuya02}). This
molecular clump has also been detected in several rotational transitions of CO
and its isotopomers, and in CS (\jtd) (Cesaroni et al.~\cite{cesa03}). In
Fig.~\ref{g24_nh_cs_ch3oh} one can see that each species peaks at different positions: \cs\ toward core A, and \METH\ and \nh\ between cores C and A, with the
\nh\ emission peaking much closer to core C. This suggests that every species
could be tracing different physical conditions and/or chemical evolution of the
molecular gas toward the YSOs. The lower resolution of our
observations, however, did not allow us to investigate this in more detail. Taking into account that we have detected
rotating structures associated with some of the individual cores embedded in
G24.78+0.08 (see Paper~I), we also searched for the presence of a global
velocity gradient that could be indicative of large
scale rotation of the whole cloud. To do this we fitted the \cs\ (\jdu) lines with a Gaussian at each point in order to measure shifts from the systemic
velocity of the cloud, and we also minimized the $\chi^2$ measure of the
difference between a model with a linear velocity gradient and the \cs\
(\jdu) data at each point. However, no significant  rotation was detected by any of
the two methods, except for the central part of the cloud where the velocity
gradient is more evident and consistent with that found on a smaller scale in \MCN.

\begin{figure}
\centerline{\includegraphics[angle=-90,width=8.5cm]{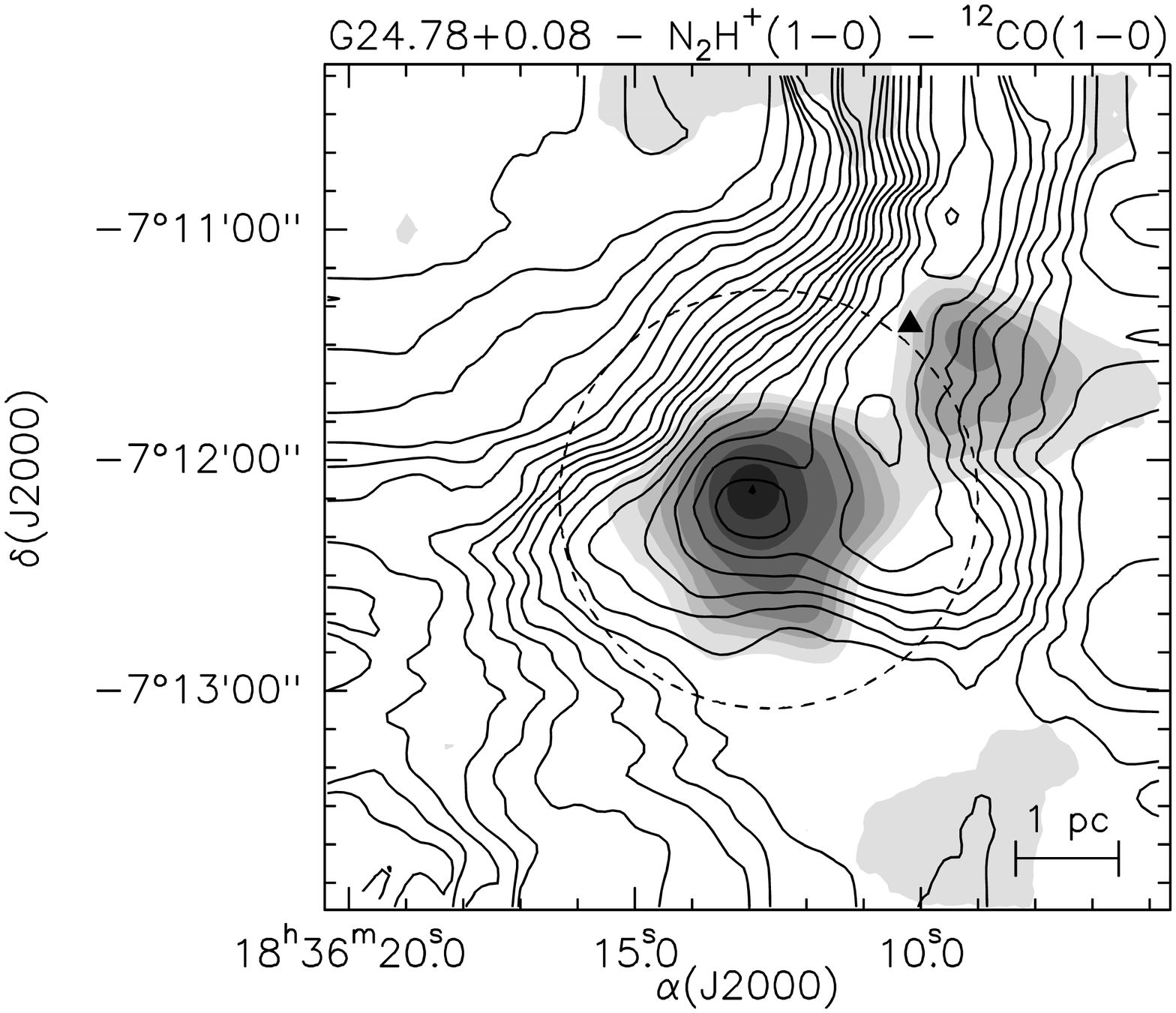}}
\caption{Overlay of the BIMA N$_2$H$^+$ (\juz) integrated emission ({\it greyscale}), and  the \CO\ (\juz) integrated emission obtained with the IRAM 30-m telescope by 
Cesaroni et al.~(\cite{cesa03}; {\it contours}) toward G24.78+0.08. The \CO\ emission has been integrated under the line, and the contour levels range from 4 to 194~K\,\kms\ in steps of 10~K\,\kms. Contours for N$_2$H$^+$ are the same as in Fig.~\ref{g24_nh_cs_ch3oh}. The triangle pinpoints the position of the peak of the 20~cm continuum emission extracted from the NVSS database. The dashed circumference represents the BIMA primary beam (50$\%$ attenuation level).}  
\label{nh_co}
\end{figure}

\begin{figure*}
\centerline{\includegraphics[angle=-90,width=11.5cm]{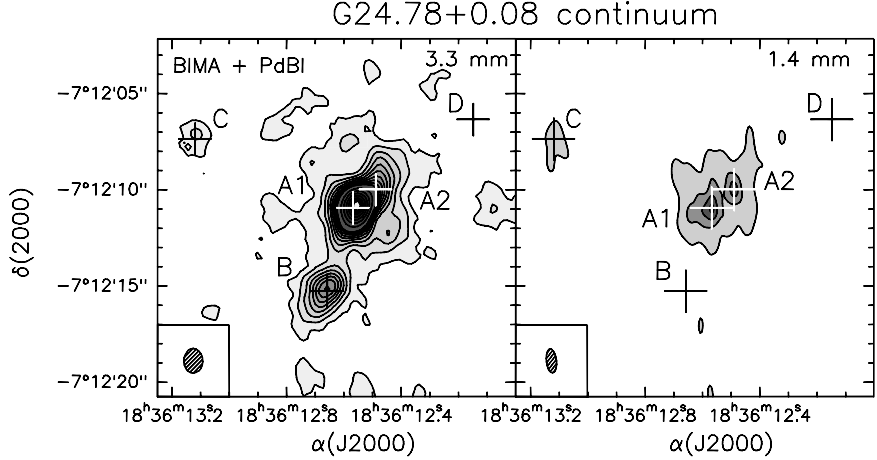}}
\vspace{0.4cm}
\caption{Combined BIMA+PdBI map of the 3.3~mm continuum emission ({\it left panel}), and PdBI map of the 1.4~mm continuum emission ({\it right panel}) continuum emission toward the core of G24.78+0.08. The contour levels range 
from 2 to 20\mjy\ in steps of 2\mjy\ and from 20 to 80\mjy\ in steps of 10\mjy\ for the 3.3~mm map, and from 20 to 200\mjy\ in steps of 60\mjy\ for the 1.4~mm map. The synthesized beam is shown in the lower left-hand corner. The crosses mark the position of the YSOs identified in the region (Furuya et al.~\cite{furuya02}; Paper I; see Sect.~\ref{g24-cont}).}
\label{g24_continuum}
\end{figure*}

A second molecular clump has clearly been detected in \nh\ and is also
visible in \METH\ and marginally in \cs, toward the northwest of G24.78+0.08 (see
Fig.~\ref{g24_nh_cs_ch3oh}). Millimeter continuum emission has also been detected in our 3.16~mm low angular resolution ($\theta \sim 18\arcsec$) BIMA observations toward the northwest. This continuum emission peaks at the position of an extended \HII region visible at 20~cm in the NVSS survey (Condon et al.~\cite{condon98}), and is probably associated with it. However, as can be seen in the figure, there is also some continuum emission associated with the \nh\ clump.
Submillimeter emission associated with this clump has
also been mapped at 850 and 450$\mu$m with the Submillimeter Common-User Bolometric
Array (SCUBA) by Walsh et al.~(\cite{walsh03}). In their maps it is also 
visible lower-level extended emission connecting the two clumps in the
region. The clump has also been
detected in several rotational transitions of CO and its isotopomers
(Cesaroni et al.~\cite{cesa03}). However, the peak of the \CO\
emission is displaced from the position of the \nh\ emission peak (see
Fig.~\ref{nh_co}), unlike the molecular clump at the center of the map for
which both CO and \nh\ peak at the same position. The lack of CO emission toward this northern clump could be due to an opacity effect or to molecular depletion. For low-mass starless cores, the abundance of some molecules, including CO and CS, dramatically decreases toward the core center, with respect to the outer region. On the other hand, molecules such as \nh\ are found to have constant abundances (e.g. Tafalla et al.~\cite{tafalla02}). These results are explained by the depletion of carbon and sulfur bearing molecules onto dust grains at high densities (a few times 10$^4$~cm$^{-3}$) and low temperatures occurring in dense core interiors (e.g.\ Bergin \& Langer~\cite{bergin97}). Nitrogen bearing molecules are instead unaffected by this process up to densities of several times 10$^5$~cm$^{-3}$, or even 10$^6$~cm$^{-3}$. In order to assess that depletion is the cause of the lack of CO emission we need to estimate the opacity and the temperature of this northern clump. However, only the CO (\juz) and (\jdu) lines have been observed toward this clump: therefore additional
observations of an optically thin CO isotopomer and of a good temperature tracer are needed. Unfortunately, the energy of the three \METH\ (\jduk)
transitions observed with BIMA are too similar to derive any temperature by means of the RD method. In case that the CO freeze-out is confirmed, the northern clump would represent the first example of a chemically depleted high-mass clump.


\subsubsection{Continuum emission}
\label{g24-cont}

In Fig.~\ref{g24_continuum} we show the maps of the combined BIMA + PdBI 3.3~mm continuum
emission and the PdBI 1.4~mm continuum emission toward G24.78+0.08. As mentioned in Sect.~\ref{bima_molecularclumps}, 3.16~mm continuum low angular resolution ($\theta \sim 18\arcsec$)  BIMA maps are shown in Fig.~\ref{g24_nh_cs_ch3oh}. The highest angular resolution ($\theta \sim 1\arcsec$) observations at 3.3~mm have resolved the continuum emission into three sources, A, B, and C, already
detected at millimeter wavelengths by Furuya et al.\ (\cite{furuya02}). The
continuum emission at 3.3~mm  is dominated by an extended structure
surrounding cores A and B. The strongest core detected at 3.3~mm wavelength in
the region is core A. The emission clearly shows two components, a centrally
peaked source, plus an extended component elongated toward the northwest. In
fact, the 1.4~mm dust emission associated
with core A has been resolved into two separate cores, cores A1 and A2 (see
Fig.~\ref{g24_continuum} and Paper I), with core A2 located in the direction of the elongation toward NW seen at
3.3~mm. The two cores are separated by $\sim 1\farcs5$ or $\sim 11500$~AU. The positions, fluxes and deconvolved diameters of the
cores are given in Table~\ref{table_cont}. The other core detected at both
wavelengths is C. The continuum emission associated with this YSO is much
less extended, and it has only been resolved at 1.4~mm. The measured flux
densities of cores A and C at 3.3~mm are consistent with the previous measurements by Furuya et al.~(\cite{furuya02}; see their Fig.~3). At 1.4~mm, the sum of
the flux densities of cores A1 and A2 is consistent with the expected flux
density for core A, while for C the flux that we have measured is slightly
lower, but still consistent, with the model
predictions of Cesaroni et al.~(\cite{cesa03}). The third source detected in the region at 3.3~mm is the UC \HII region B (Codella et al.~\cite{codella97}; see
Fig~\ref{g24_continuum}), which is slightly elongated in the southeast-northwest
direction. This has not been detected at 1.4~mm; however, the 1.4~mm flux expected on the basis of the results of Cesaroni
et al.~(\cite{cesa03}) is lower than
the $3\sigma$ noise level of our map, which is about 20~\mjy. The flux
at 3.3~mm is also consistent with the values of Cesaroni et
al.~(\cite{cesa03}). Finally, we have not detected at any wavelength core
D, which has been detected at 2.6 and 2~mm by Furuya et al.~(\cite{furuya02}).
However, the expected flux of this core at 3.3 and 1.4~mm, according to the
predictions of Cesaroni et al.~(\cite{cesa03}), falls below the $3\sigma$ noise
level of our maps, 2 and 20~mJy, respectively.

\begin{figure}
\centerline{\includegraphics[angle=0,width=7.5cm]{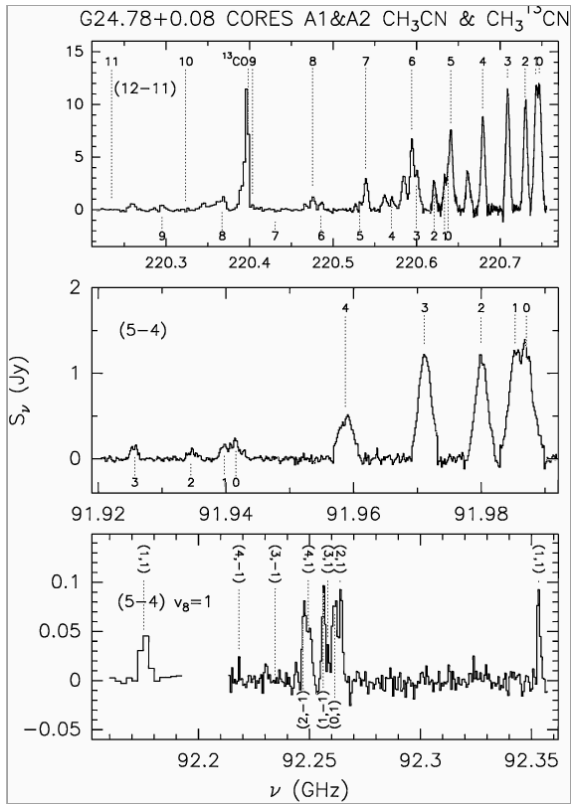}}
\caption{Same as Fig.~\ref{g31-spectra} for cores A1+A2 in G24.78+0.08.} 
\label{g24-spectra}
\end{figure}

\begin{figure}
\centerline{\includegraphics[angle=0,width=7.5cm]{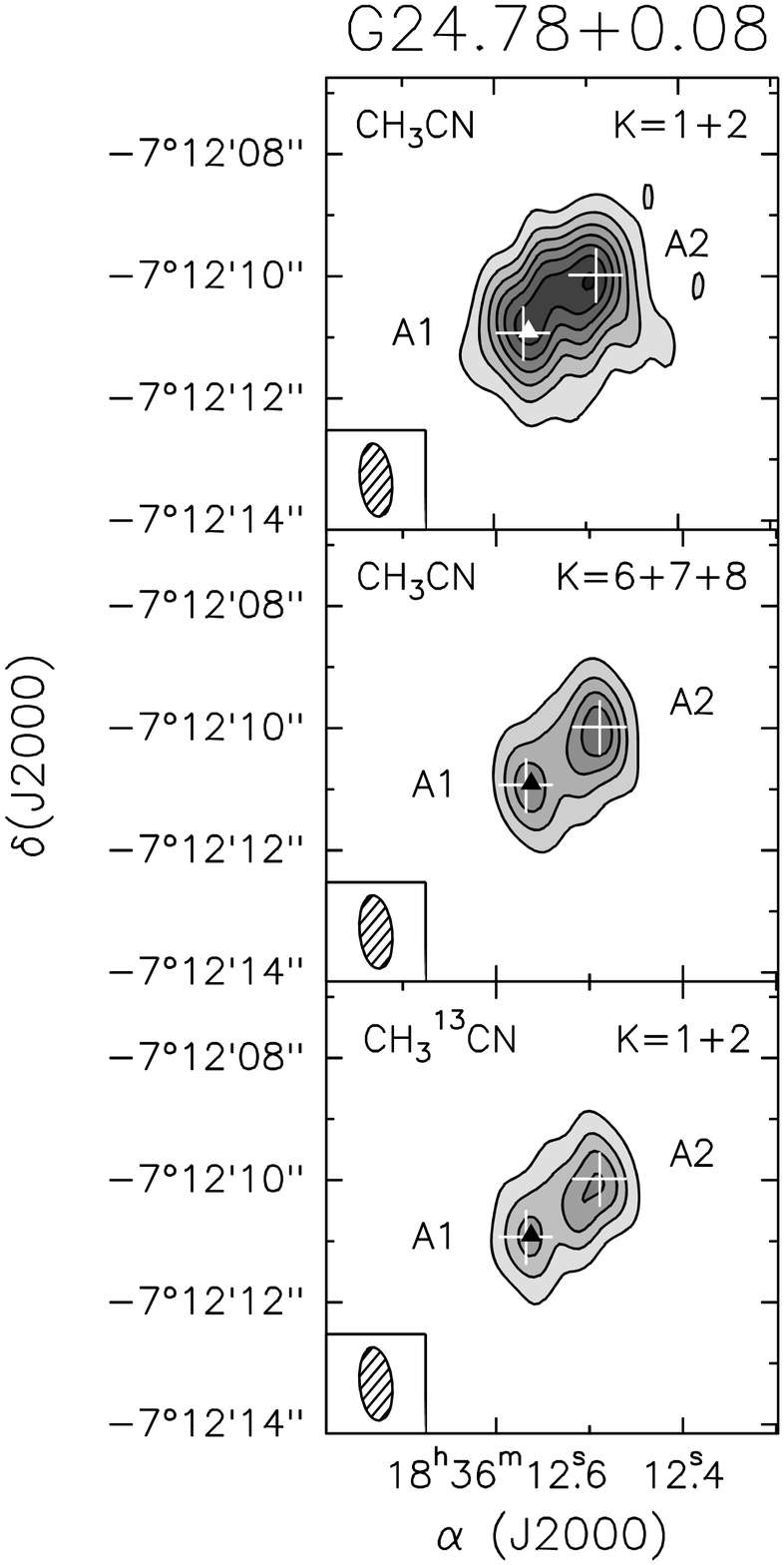}}
\caption{PdBI intensity maps of the \MCN\ (\jdo) emission averaged under the $K=1$ and 2 components ({\it top panel}), under the $K=6$, 7, and 8 components ({\it middle panel}), and of the \MCNII\ (\jdo) emission under the $K=1$ and 2 components ({\it bottom panel}) toward cores A1 and A2 in G24.78+0.08. The contour levels range from 0.14 to 0.98\jy\ ({\it top panel}), and to 0.56\jy\ ({\it middle} and {\it bottom panels}), in steps of 0.14\jy. The synthesized beam is shown in the lower left-hand corner.
The white crosses denote the position of the 1.4~mm continuum emission peaks. The triangle marks the position of the VLA 1.3~cm continuum emission peak (Codella et al.~\cite{codella97}).}
\label{g24_ch3cn_1mm}
\end{figure}

\begin{figure}
\centerline{\includegraphics[angle=0,width=7.5cm]{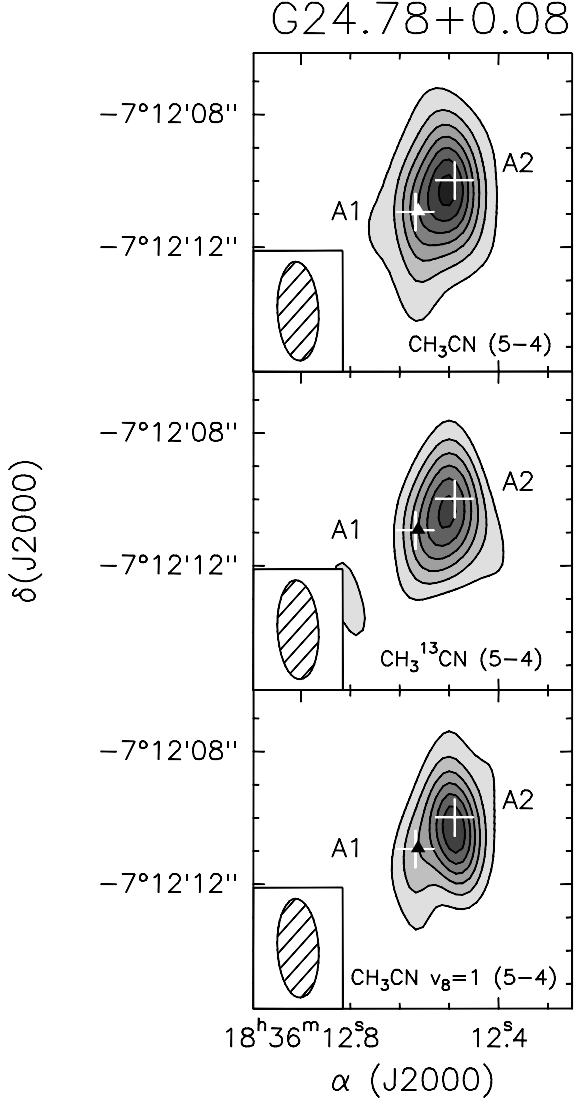}}
\caption{PdBI intensity maps of the \MCN\ (\jcc) ({\it top panel}) emission averaged under the $K=0$ to 4 components, the \MCNII\ ({\it middle panel}) emission averaged under the $K=0$ to 3 components, and of the $v_8=1$ emission ({\it bottom panel}) averaged under the (1,1), (2,1), (0,1), (3,1), (1,$-1$), (4,1), (2,$-1$), (3,$-1$), (4,$-1$), and (1,1) transitions toward cores A1 and A2 in G24.78+0.08. The contour levels range 
from 0.05 to 0.35\jy\ in steps of 0.05\jy\ ({\it top panel}), from 0.01 to 0.06\jy\ in steps of 0.01\jy ({\it middle panel}), and from 
0.005 to 0.030\jy\ in steps of 0.005\jy ({\it bottom panel}). The synthesized beam is shown in the lower left-hand corner.
The white crosses denote the position of the 1.4~mm continuum emission peaks. The triangle marks the position of the VLA 1.3~cm continuum emission peak (Codella et al.~\cite{codella97}).}
\label{g24_ch3cn_3mm}
\end{figure}

\subsubsection{\MCN\ and \MCNII}
\label{g24-ch3cn}

PdBI spectra of the \MCN\ (\jdo) and (\jcc), \MCNII\ (\jdo) and (\jcc), and
\MCN\ (\jcc) $v_8=1$ lines toward cores A1 and A2 in G24.78+0.08 are shown in
Fig.~\ref{g24-spectra}.  The spectra have been obtained averaging the emission
over the $3\sigma$ contour level area, which surrounds both cores A1 and A2.
Figure~\ref{g24_ch3cn_1mm} shows the PdBI maps of the \MCN\ (\jdo) emission
averaged  under the $K=$1 and 2 components, and under the $K=6$, 7 and 8
components, and the \MCNII\ emission averaged under the $K=$1 and 2 components.
Figure~\ref{g24_ch3cn_3mm} shows the maps of the \MCN\ (\jcc) emission averaged
under the $K=0$ to 4 components, the \MCNII\ emission averaged under the $K=0$
to 3 components, and the \MCN\ $v_8=1$  emission averaged under all observed
transitions.

The methyl cyanide
emission of the two cores has not been resolved at 3.3~mm because the
separation between them is smaller than the synthesized beam, whereas they have been resolved at 1.4~mm (Figs.~\ref{g24_ch3cn_1mm} and \ref{g24_ch3cn_3mm}). For the sake of comparison with the \MCN\ (\jcc) lines, also for the \MCN\ (\jdo) emission we decided to plot the spectra integrated over a region surrounding both cores.  As can be
seen in Fig.~\ref{g24-spectra} several $K$-components of the different 
rotational
transitions of methyl cyanide were clearly detected. No methyl cyanide emission
has been detected toward cores B or D with the PdBI. For core C, only  the \MCN\
(\jcc) emission has been barely detected, with an averaged intensity peak of $\sim
30$\mjy. Cores B and C have been detected in the \MCN\ (8--7) transition by Furuya et al.~(\cite{furuya02}), hence the fact that they are not detected in the (\jdo) lines could be due to these
cores being significantly colder than cores A1 and A2. As done for the
spectra toward G31.41+0.31, we fitted the multiple $K$-components of the
different \MCN\ and \MCNII\ rotational transitions (see Sect.~\ref{g31-ch3cn}
for details on the fitting).  The \MCN\ and \MCNII\ (\jdo) line parameters for
cores A1 and A2, individually, are given in Table~\ref{table_ch3cn_a1} and
\ref{table_ch3cn_a2}, respectively, while the \MCN\ and \MCNII\ (\jcc),
and \MCN\ (\jcc) $v_8=1$ line parameters for the emission surrounding both
cores A1 and A2 (hereafter A1+A2) are given in Table~\ref{table_ch3cn_a} and \ref{table_v8_a}.  As
can be seen in Tables~\ref{table_ch3cn_a1} and \ref{table_ch3cn_a2} the systemic velocities and
the line widths derived for the (\jdo) transition are quite similar for both
cores: \Vlsr~$\simeq 110.6$--110.9~\kms, and FWHM~$\simeq 7$~\kms. The
systemic velocity is also consistent with that found for the ground state and  $v_8=1$ (\jcc) transitions in cores A1+A2, although those
derived from the ground state lines are slightly higher (111.2--111.6~\kms). Regarding the line
widths, the values derived for the (\jcc) transitions are quite different, with $\sim 7.8$~\kms\ for \MCN\ and  $\sim 4.6$~\kms\ for \MCNII. As done for G31.41+0.31 (see Sect.~\ref{g31-ch3cn}), one can derive the ratio between the observed and the intrinsic line widths, which for G24.78+0.08 is $1.7\pm0.04$. On the other hand, the ratio derived from the optical depths of the \MCN\ and \MCNII\ (\jcc) lines (assuming a relative abundance [\MCN]/[\MCNII]=40; see Wilson \& Rood~\cite{wilson94}) ranges from 1.5 to 1.9. Thus, also in this case, optical depth effects could be responsible for the variation of the line widths. The line widths found for (\jcc) $v_8=1$ are slightly
smaller but still consistent with those found for the (\jdo) transition.

\begin{figure}
\centerline{\includegraphics[angle=0,width=7.5cm]{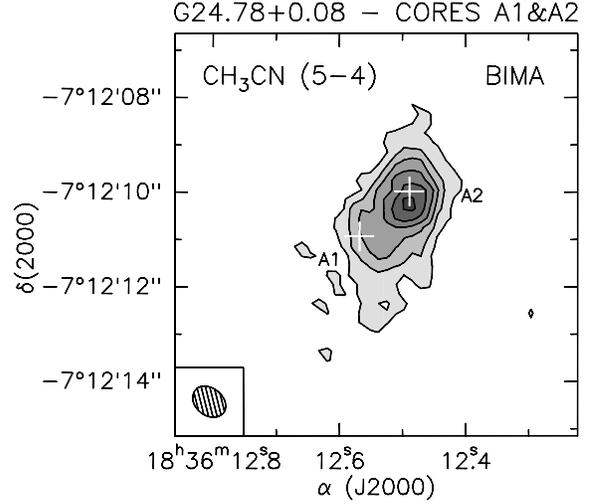}}
\caption{BIMA integrated intensity map of the \MCN\ (\jcc) emission under the $K=0$, 1, 2, and 3 components toward cores A1 and A2 in G24.78+0.08. The contour levels are 3, 6, 9, 12, 15, and 18 times 0.35\jykms. The synthesized beam is shown in the lower left-hand corner. The white crosses mark the position of the 1.4~mm continuum emission peaks.}
\label{g24_ch3cn_1_bima}
\end{figure}

\begin{figure}
\centerline{\includegraphics[angle=0,width=7.5cm]{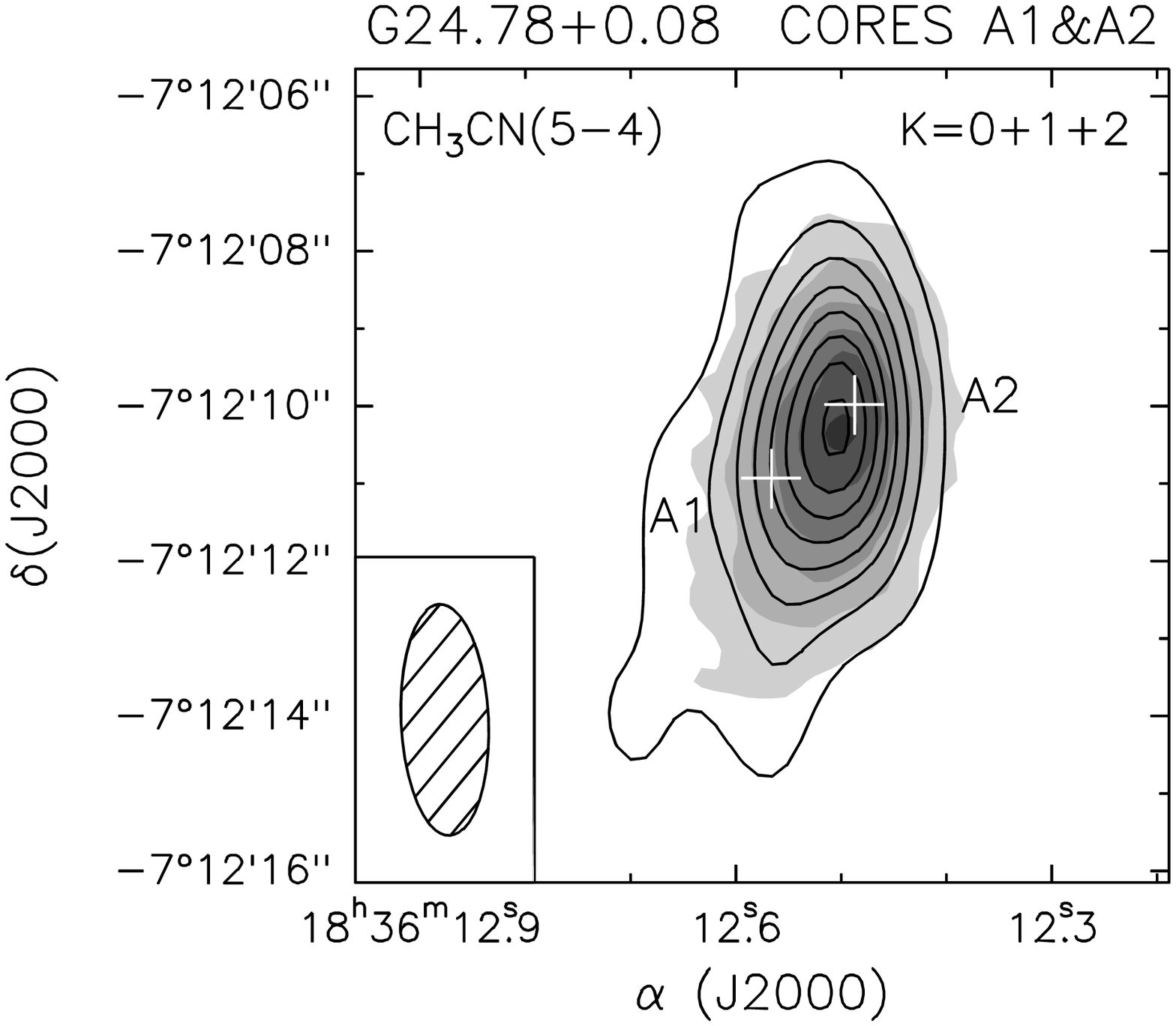}}
\caption{Overlay of the integrated intensity map of the \MCN\ (\jcc) emission under the $K=0$, 1, and 2 components obtained with the PdBI ({\it contours}) on the map obtained with BIMA ({\it greyscale}). The contour levels range 
from 0.05 to 0.35\jykms\ ({\it greyscale}) and to 0.40\jykms\ ({\it contours}) in steps of 0.05\jykms. The BIMA  map has been obtained by restoring the 'clean' map with the same beam than that of the PdBI observations (see text). The synthesized beam is shown in the lower left-hand corner.The white crosses mark the position of the 1.4~mm continuum emission peaks.}
\label{ch3cn_mom0_bima_pdbi}
\end{figure}

\begin{figure}
\centerline{\includegraphics[angle=270,width=7.5cm]{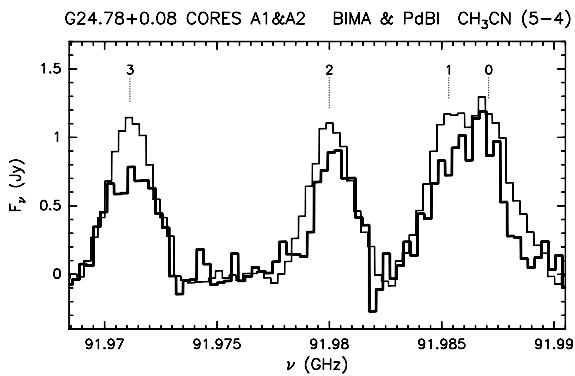}}
\caption{Comparison between the BIMA (thick line) and the PdBI (thin line) \MCN\ (\jcc) spectra toward cores A1+A2 in G24.78+0.08. Both spectra have been obtained averaging the emission over the same region. The BIMA channel maps have been obtained by restoring the 'clean' map with the same beam than that of the PdBI observations (see text). The  number of the different $K$-components are marked with dashed lines in the upper part.} 
\label{ch3cn_54_bima_pdbi}
\end{figure}

 Both core A1 and A2 are clearly visible in Fig.~\ref{g24_ch3cn_1mm}, being the emission
associated with core A2 the strongest for all the transitions. The emission of both cores has not been resolved at 3~mm (Fig.~\ref{g24_ch3cn_3mm}). As can be seen in Fig.~\ref{g24_ch3cn_1mm}, the morphology of the emission is quite similar for the
different $K$-components of \MCN\ and \MCNII: one sees two compact cores,
roughly coincident with the two 1.4~mm continuum cores A1 and A2, plus an
extended component. Such a good agreement between the gas and the dust emission indicates that
they are tracing the same material. In addition, there is no clear evidence of
any emission dip for lower $K$-components of \MCN\ toward the position of the
embedded sources in the two cores, as for G31.41+0.31. This
suggests that the temperature gradient in
cores A1 and A2 in G24.78+0.08 is less prominent than in G31.41+0.31. This is confirmed by the $T_{\rm rot}$ measured in each core, as illustrated in Sect.~\ref{g24_temp}. In Fig.~\ref{g24_ch3cn_1mm} we also mark the position of the peak of the 1.3~cm continuum emission of the UC \HII region detected with the VLA by Codella et al.~(\cite{codella97}). Note that this \HII region is associated with core A1, being located at $\sim 0\farcs1$ ($\sim 800$~AU at the distance of the region) of the millimeter continuum source, whereas no free-free emission is reported toward core A2.

\begin{table*}
\caption[] {\MCN\ and \MCNII\ (\jdo) line parameters for core A1 in G24.78+0.08}
\label{table_ch3cn_a1}
\begin{tabular}{lccccc}
\hline
\multicolumn{1}{c}{Line} &
\multicolumn{1}{c}{$V_{\rm LSR}$} &
\multicolumn{1}{c}{FWHM} &
\multicolumn{1}{c}{$T_{\rm B}^{\rm a}$} &
\multicolumn{1}{c}{$\int{T_{\rm B}\,{\rm d}V}$} &
\multicolumn{1}{c}{rms} \\
\multicolumn{1}{c}{$K$} &
\multicolumn{1}{c}{(\kms)} &
\multicolumn{1}{c}{(\kms)} &
\multicolumn{1}{c}{(K)} & 
\multicolumn{1}{c}{(K \kms)} &
\multicolumn{1}{c}{(K \kms)} \\
\hline
&&&\MCN\   \\
\hline
0 &110.82$\pm0.02$ &$7.25\pm0.03$ &$26.8\pm1.5$ &$207\pm3$ &0.9\\
1 &" &" &$27.6\pm1.5$ &$213\pm3$ &"\\
2 &" &" &$28.3\pm1.5$ &$219\pm3$ &" \\
3 &" &" &$30.3\pm1.5$ &$234\pm2$ &"\\
4 &110.82$\pm0.05$ &$7.12\pm0.07$ &$23.6\pm1.3$ &$179\pm3$ &1.0\\
5 &" &" &$18.2\pm1.3$ &$207\pm3$ &"\\
6 &$110.64\pm0.12$ &$6.90\pm0.16$ &$17.0\pm0.8$ &$125\pm5$ &0.8 \\
7 &" &" &$7.7\pm0.8$ &$56\pm3$ &0.8 \\
8 &" &" &$3.4\pm0.8$ &$25\pm3$ &" \\
9 &" &" &(b) &(b) &" \\
10 &" &" &$<0.8$ & &" \\
11 &" &" &$<0.8$ & &" \\
\hline
&&&\MCNII\   \\
\hline
0 &110.82$\pm0.05$ &$7.12\pm0.07$ &(c) &(c) &1.0\\
1 &" &" &$7.0\pm1.3$ &$53\pm3$ &"\\
2 &" &" &$6.6\pm1.3$ &$50\pm2$ &" \\
3 &$110.64\pm0.12$ &$6.90\pm0.16$ &$8.7\pm0.8$ &$64\pm5$ &0.8 \\
4 &$110.64\pm0.10$ &$6.92\pm0.14$ &$2.6\pm0.8$ &$19\pm3$ &0.7 \\
5 &" &" &$0.9\pm0.8^{\rm d}$ &$6\pm3^{\rm d}$ &" \\
6 &$110.64\pm0.12$ &$6.90\pm0.16$ &$1.8\pm0.8$ &$13\pm3$ &0.8 \\
7 &$110.68\pm0.72$ &$7.0^{\rm e}$ &$<0.8$ & &0.8 \\
8 &" &" &$3.4\pm0.5^{\rm f}$ &$26\pm5^{\rm f}$ &" \\
9 &" &" &$1.3\pm0.5$ &$10\pm5$ &" \\
\hline

\end{tabular}
  
  (a) Brightness temperature integrated over a $3\sigma$ emission level area around core A1. \\
  (b) Blended with $^{13}$CO (\jdu). \\
  (c) Blended with \MCN\ $K=5$. \\
  (d) Blended with \MCN\ $K=7$. \\
  (e) This parameter was held fixed in the fit. \\
  (f) Possibly blended with wing emission of  $^{13}$CO (\jdu).   
  
\end{table*}

\begin{table*}
\caption[] {\MCN\ and \MCNII\ (\jdo) line parameters for core A2 in G24.78+0.08}
\label{table_ch3cn_a2}
\begin{tabular}{lccccc}
\hline
\multicolumn{1}{c}{Line} &
\multicolumn{1}{c}{$V_{\rm LSR}$} &
\multicolumn{1}{c}{FWHM} &
\multicolumn{1}{c}{$T_{\rm B}^{\rm a}$} &
\multicolumn{1}{c}{$\int{T_{\rm B}\,{\rm d}V}$} &
\multicolumn{1}{c}{rms} \\
\multicolumn{1}{c}{$K$} &
\multicolumn{1}{c}{(\kms)} &
\multicolumn{1}{c}{(\kms)} &
\multicolumn{1}{c}{(K)} & 
\multicolumn{1}{c}{(K \kms)} &
\multicolumn{1}{c}{(K \kms)} \\
\hline
&&&\MCN\   \\
\hline
0 &110.68$\pm0.03$ &$7.25\pm0.03$ &$22.0\pm1.6$ &$170\pm3$ &1.1\\
1 &" &" &$22.6\pm1.6$ &$174\pm3$ &"\\
2 &" &" &$23.8\pm1.6$ &$184\pm3$ &" \\
3 &" &" &$26.2\pm1.6$ &$202\pm2$ &"\\
4 &110.59$\pm0.07$ &$7.18\pm0.10$ &$20.8\pm0.9$ &$159\pm3$ &1.3\\
5 &" &" &$16.0\pm0.9$ &$123\pm5$ &"\\
6 &$110.87\pm0.08$ &$6.74\pm0.11$ &$16.3\pm1.1$ &$117\pm3$ &0.6 \\
7 &" &" &$7.9\pm1.1$ &$57\pm2$ &" \\
8 &$110.88\pm0.10$ &$6.66\pm0.13$ &$4.3\pm1.0$ &$30\pm3$ &0.7 \\
9  &$110.50\pm0.60$ &$7.0^{\rm b}$ &(c) &(c) &0.8 \\
10 &" &" &$<0.8$ & &" \\
11 &" &" &$<0.8$ & &" \\
\hline
&&&\MCNII\   \\
\hline
0 &110.59$\pm0.07$ &$7.18\pm0.10$ &(d) &(d) &1.3\\
1 &" &" &$6.1\pm0.9$ &$47\pm3$ &"\\
2 &" &" &$6.0\pm0.9$ &$46\pm3$ &" \\
3 &$110.87\pm0.08$ &$6.74\pm0.11$ &$9.3\pm1.1$ &$67\pm3$ &0.6 \\
4 &" &" &$3.3\pm1.1$ &$24\pm2$ &" \\
5 &" &" &$1.8\pm1.1^{\rm e}$ &$13\pm3^{\rm e}$ &" \\
6 &$110.88\pm0.10$ &$6.66\pm0.13$ &$2.1\pm1.0$ &$15\pm3$ &0.7 \\
7 &$110.50\pm0.60$ &$7.0^{\rm b}$ &$<0.8$ & &0.8 \\
8 &" &" &$3.7\pm0.6^{\rm f}$ &$27\pm5^{\rm f}$ &" \\
9 &" &" &$1.2\pm0.6$  &$9\pm5$ &" \\
\hline

\end{tabular}

  (a) Brightness temperature integrated over a $3\sigma$ emission level area around core A2. \\
  (b) This parameter was held fixed in the fit. \\
  (c) Blended with $^{13}$CO (\jdu). \\
  (d) Blended with \MCN\ $K=5$. \\
  (e) Blended with \MCN\ $K=7$. \\
  (f) Dominated by wing emission of  $^{13}$CO (\jdu).   
  
\end{table*}

\begin{table*}
\caption[] {\MCN\ and \MCNII\ (\jcc) line parameters for core A1+A2 in G24.78+0.08$^{\rm a}$}
\label{table_ch3cn_a}
\begin{tabular}{lccccc}
\hline
\multicolumn{1}{c}{Line} &
\multicolumn{1}{c}{$V_{\rm LSR}$} &
\multicolumn{1}{c}{FWHM} &
\multicolumn{1}{c}{$T_{\rm B}^{\rm b}$} &
\multicolumn{1}{c}{$\int{T_{\rm B}\,{\rm d}V}$} &
\multicolumn{1}{c}{rms} \\
\multicolumn{1}{c}{$K$} &
\multicolumn{1}{c}{(\kms)} &
\multicolumn{1}{c}{(\kms)} &
\multicolumn{1}{c}{(K)} & 
\multicolumn{1}{c}{(K \kms)} &
\multicolumn{1}{c}{(K \kms)} \\
\hline
&&&\MCN\ \\
\hline
0 &$111.61\pm0.02$ &$7.74\pm0.03$ &$10.6\pm1.0$ &$87\pm1$ &0.3\\
1 &" &" &$8.9\pm1.0$ &$70\pm1$ &" \\
2 &" &" &$10.8\pm1.0$ &$89\pm1$ &" \\
3 &" &" &$11.6\pm1.0$ &$96\pm1$ &" \\
4 &" &" &$4.9\pm1.0$ &$40\pm1$ &" \\
\hline
&&&\MCNII\   \\
\hline
0 &$111.20\pm0.08$ &$4.58\pm0.09$ &$2.00\pm0.27$ &$9.8\pm0.4$ &0.18\\
1 &" &" &$1.51\pm0.27$ &$7.39\pm0.35$  &" \\
2 &" &" &$0.86\pm0.27$ &$4.21\pm0.35$ &" \\
3 &" &" &$1.33\pm0.27$ &$6.50\pm0.35$ &" \\
\hline

\end{tabular}

  (a) Emission integrated over a region surrounding cores A1 and A2 in G24.78+0.08.\\
  (b) Brightness temperature integrated over a $3\sigma$ emission level area. \\
\end{table*}

\begin{table*}
\caption[] {\MCN\  (\jcc) $v_8=1$ line parameters for core A1+A2 in G24.78+0.08$^{\rm a}$}
\label{table_v8_a}
\begin{tabular}{lccccc}
\hline
\multicolumn{1}{c}{Line} &
\multicolumn{1}{c}{$V_{\rm LSR}$} &
\multicolumn{1}{c}{FWHM} &
\multicolumn{1}{c}{$T_{\rm B}^{\rm b}$} &
\multicolumn{1}{c}{$\int{T_{\rm B}\,{\rm d}V}$} &
\multicolumn{1}{c}{rms} \\
&\multicolumn{1}{c}{(\kms)} &
\multicolumn{1}{c}{(\kms)} &
\multicolumn{1}{c}{(K)} & 
\multicolumn{1}{c}{(K \kms)} &
\multicolumn{1}{c}{(K \kms)} \\
\hline
(1,1) &$110.61\pm0.18$ &$6.26\pm0.19$ &$0.77\pm0.14$ &$5.15\pm0.34$ &0.07\\
(2,1) &$110.84\pm0.12$ &$5.78\pm0.13$ &$0.82\pm0.14$ &$5.04\pm0.32$ &0.07\\
(0,1) &" &" &$0.91\pm0.14$ &$5.61\pm0.32$ &" \\
(3,1) &" &" &$0.22\pm0.14$ &$1.38\pm0.31$ &"\\
$(1,-1)$ &" &" &$0.83\pm0.14$ &$5.11\pm0.32$ &" \\
$(4,1)$ &$110.61\pm0.18$ &$6.26\pm0.19$ &$0.60\pm0.14$ &$3.98\pm0.34$ &0.07 \\
$(2,-1)$ &" &" &$0.73\pm0.14$ &$4.84\pm0.34$ &"\\
$(3,-1)$ &" &" &$<0.07$ & &" \\
$(4,-1)$ &$110.61\pm0.18$ &$6.26\pm0.19$ &$0.11\pm0.14$ &$0.71\pm0.32$ &0.07 \\
$(1,1)$ &$110.61^{\rm c}$ &$6.26^{\rm c}$ &$0.89\pm0.09^{\rm d}$ &$5.94\pm0.52^{\rm e}$ &0.04 \\
\hline

\end{tabular}

 (a) Emission integrated over a region surrounding G24.78+0.08 cores A1 and A2.\\
 (b) Brightness temperature integrated over a $3\sigma$ emission level area. \\
 (c) This parameter was held fixed in the fit. \\
 (d) Computed from the integrated line intensity. \\  
 (e) Measured with the command print area of CLASS. \\  

\end{table*}

Figure~\ref{g24_ch3cn_1_bima} shows the BIMA integrated \MCN\ (\jcc) emission under the
$K=0$, 1, 2, and 3 components toward cores A1 and A2. The cores have been clearly resolved in these observations
($\theta \sim 0\farcs7$). As can be seen in the figure the strongest emission
comes from core A2, with the peak slightly displaced from that
of the 1.4~mm continuum emission. On the other hand, the emission from core A1
is hardly detected.  This could be due to a temperature effect, being core A1 probably hotter than core A2. We checked whether it could be an opacity effect by computing the optical depth in the cores from the ratio of \MCNII\ and \MCN\ (\jdo) $K=2$ component (see Sect.~\ref{g31-ch3cn}). However, the optical depths obtained are similar for both cores, $\tau_{12}\sim 12$, suggesting that  opacity cannot account for the difference in \MCN\ (\jcc) emission between both cores. In Fig.~\ref{ch3cn_mom0_bima_pdbi} we compare the \MCN\ (\jcc) maps obtained with BIMA and the PdBI. To compare them we have first reduced the
resolution of the BIMA map by restoring the 'clean' BIMA map with the same beam
as the PdBI observations. As can be seen in
this figure there is an excellent agreement between the two maps, which confirms
that most of the \MCN\ (\jcc) emission is associated with core A2. Note that
some extended emission is not shown by the BIMA map, probably due to the fact
that it is filtered out by the interferometer due to the lack of short spacings.  In Fig.~\ref{ch3cn_54_bima_pdbi} the same comparison is shown for the spectra obtained integrating the emission over the same region. Again, taking into account that some of the emission is probably filtered out by the BIMA interferometer, one can say that the agreement between both spectra is remarkable.


\subsection{Discussion}

\subsubsection{Temperature and column density}
\label{g24_temp}

\begin{figure}
\centerline{\includegraphics[angle=-90,width=8cm]{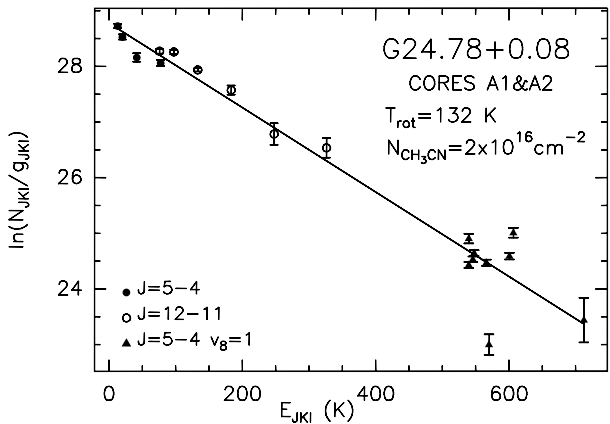}}
\caption{Same as Fig.~\ref{g31-boltz} for the cores A1+A2 in G24.78+0.08.}
\label{g24-boltz}
\end{figure}

\begin{figure}
\centerline{\includegraphics[angle=-90,width=8cm]{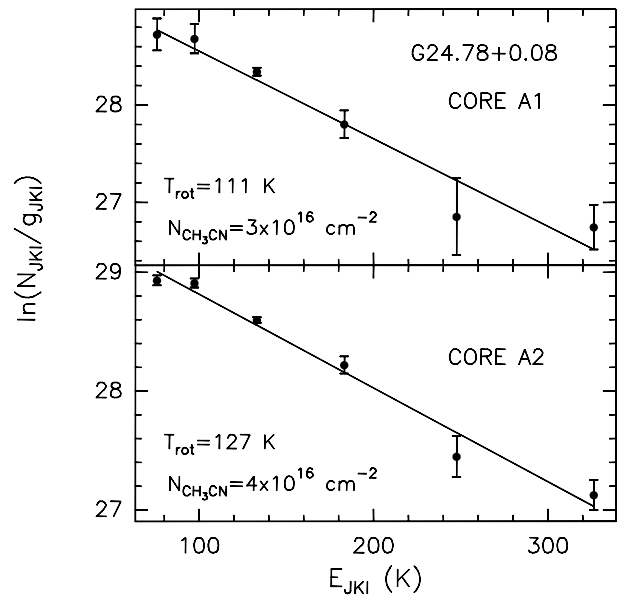}}
\caption{Rotation diagram of \MCNII\ (\jdo) for cores A1 ({\it top}) and A2 ({\it bottom}) in G24.78+0.08, with superimposed fits.}
\label{g24_a1_a2_trot}
\end{figure}

We derived the rotational temperature and total column density of the methyl
cyanide molecules by means of the RD method (see Sect.~\ref{g31-ch3cn}) for cores A1+A2 in G24.78+0.08 (see Fig.~\ref{g24-boltz}). The insufficient angular resolution of the 3~mm observations makes unable to separate the emission from the two cores, and thus the emission has been integrated over an area of $\sim 4\arcsec$ of
diameter ($\sim 12~{\rm arcsec}^2$) that covers both core A1 and A2. As 
stated before, such a region would
correspond to core A of Codella et al.~(\cite{codella97}) and  Furuya et
al.~(\cite{furuya02}). As already done for G31.41+0.31, we used only the ground
state transitions of the isotopomer \MCNII, and those from the \MCN\ $v_8=1$
state in the fit performed to the Boltzmann plot, because the ground level
transitions (see Fig.~\ref{g24-spectra}) appear to be optically thick. We assumed a
relative abundance [\MCN]/[\MCNII]=40 (see Wilson \& Rood
\cite{wilson94}). The $T_{\rm rot}$ and source averaged $N_{\rm CH_3CN}$ that
we derived by taking into account all transitions are 132~K and
$2\times10^{16}$~\cmq, respectively. The $T_{\rm rot}$ and $N_{\rm CH_3CN}$
that we estimated from each set of \mbox{$J+1\rightarrow J$} transitions of \MCNII\ are $102\pm5$ and $130\pm2$~K,
and $1\times10^{16}$ and $2\times10^{16}$~\cmq\, for the (\jcc) and (\jdo)
transitions, respectively. Note that unlike the case of G31.41+0.31, for G24.78+0.08 these two estimates are quite similar, which could be indicating that all the transitions are tracing more or less the same gas or, as mentioned before, that the temperature gradient is less evident. From NH$_3$ emission averaged over a similar area Codella et al.~(\cite{codella97}) have derived a slightly lower kinetic temperature, $T_{\rm k}=87$~K: this is not surprising because NH$_3$ is probably tracing a different region, somewhat more extended and colder. The estimate of the gas column density derived by these authors is consistent with ours. Taking into account that with the angular resolution achieved at 1~mm it is possible to separate the emission from cores A1 and A2, we made rotation diagrams for the \MCNII\ (\jdo) for each of them separately (see Fig.~\ref{g24_a1_a2_trot}). As seen in this plot, the $T_{\rm rot}$ and source averaged  $N_{\rm CH_3CN}$ are similar for both cores.

\begin{figure}
\centerline{\includegraphics[angle=0,width=8.5cm]{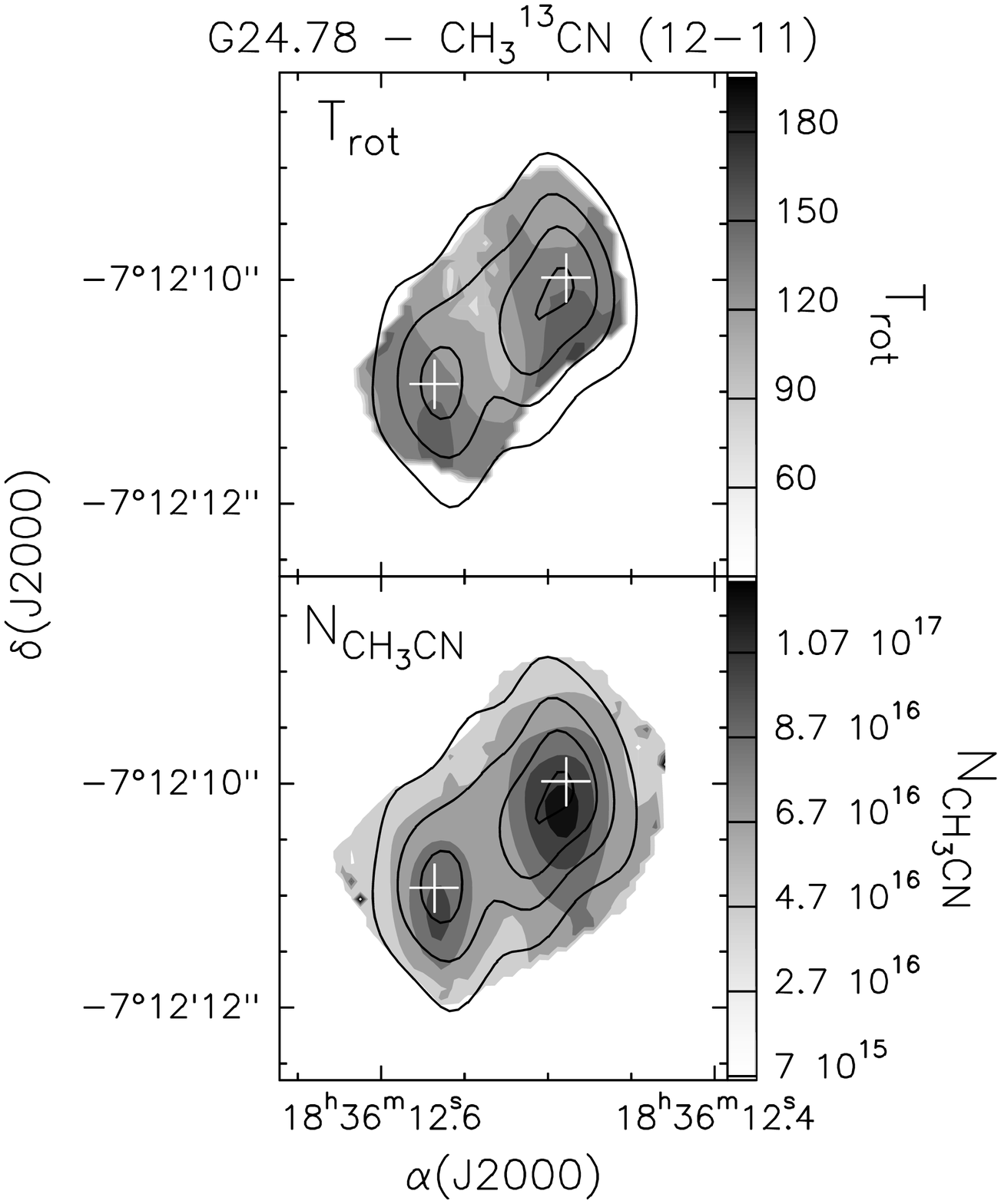}}
\caption{Overlay of the PdBI \MCNII\ (\jdo) emission averaged under the $K=1$ and 2 components ({\it contours}) on the $T_{\rm rot}$ map ({\it top panel}) and the $N_{\rm CH_3CN}$ map ({\it bottom panel}) in greyscale, derived by fitting  the Boltzmann plots at each position, toward cores A1 and A2 in G24.78+0.08. The contour levels range from 0.14 to 0.56\jy\ in steps of 0.56\jy. Greyscale levels range from 60 to 180~K by 30~K ({\it top panel}), and from $7\times 10^{15}$ to  by $1.7\times 10^{17}$~\cmq\ times by $2\times 10^{16}$~\cmq\ ({\it bottom panel}). The white crosses mark the position of the 1.4~mm continuum emission peaks.}
\label{g24_trot_ntot}
\end{figure}

In Fig.~\ref{g24_trot_ntot} we show the $T_{\rm rot}$ map and the $N_{\rm CH_3CN}$
map overlaid to the \MCNII\ (\jdo) emission averaged under the $K=$1 and 2
components toward cores A1 and A2. We computed the maps as done for G31.41+0.31
(see Sect.~\ref{g31-ch3cn}). This figure shows that both cores have similar
$T_{\rm  rot}$, $\sim120$--150~K, consistent with the source averaged
temperatures derived in the previous paragraph. In Sect.~\ref{g24-ch3cn} we proposed that core A1 could be hotter than A2, because the former is hardly detected in \MCN\ (\jcc) (see
Figs.~\ref{g24_ch3cn_1_bima} and \ref{ch3cn_mom0_bima_pdbi}). The fact
that we derive similar temperatures for both cores by means of the RD method rules out this hypothesis and indicates that the assumption of LTE is not valid in this case. In order to explain the difference in \MCN\ (\jcc) emission, one should take into account the IR radiation field of the UC \HII region
associated with this core (see Fig.~\ref{g24_ch3cn_1mm}).
Figure~\ref{g24_trot_ntot} also shows that there is no clear radial temperature
variation in the cores, unlike for G31.41+0.31 (see
Fig.~\ref{g31_trot_ntot}). Regarding the column density map, $N_{\rm CH_3CN}$, one
can see that there are two maxima roughly at the position of the methyl cyanide
and millimeter continuum emission peaks for cores A1 and A2
(Fig.\ref{g24_trot_ntot}), confirming that for these two cores, \MCN\ and dust
trace the same material.

\subsubsection{Nature of core A and evolutionary sequence}
\label{nature_A}

Core A was first detected at 1.3~cm by Codella et al.~(\cite{codella97}) as a very
compact and dense unresolved NH$_3$ core. The  gas
and dust millimeter emission from this core is the strongest and most extended
of all the cores detected in the region. In addition, Furuya et
al.~(\cite{furuya02}) have mapped a bipolar molecular outflow
that is probably powered by a YSO embedded in the core. As already mentioned, thanks to the higher
angular resolution of our PdBI millimeter observations, we have resolved 
core A into two separate cores, A1 and A2 (Paper I). These have been
detected in gas and dust emission and both could be hosting high-mass YSOs. In
such a case, the YSOs might be in different evolutionary stages, because
the two cores hosting them have roughly the same mass ($\sim 130~M_\odot$ for A1 and $\sim 80~M_\odot$ for A2; see Paper~I), but only A1 has been found in association with an UC \HII region, i.e.\  a more evolved YSO (see Fig.~\ref{g24_ch3cn_1mm}). The UC \HII region has a diameter (at
50\% of the peak emission) of $\sim 0\farcs15$ or $\sim 1200$~AU and is hence very small and probably in an early stage of its evolution. Regarding the
bipolar outflow detected by Furuya et
al.~(\cite{furuya02}) in association with A, it is not evident which of the cores, A1
or A2, hosts the powering source. Core A1 lies slightly closer to the
geometrical center of the outflow,
but the small separation between the cores and the fact that they are
aligned along the outflow axis make it difficult to establish any association unambiguously. Sensitive mid-IR observations would be very helpful to assess the YSO content of each core and possibly constrain their evolutionary state.

Furuya et al.~(\cite{furuya02}) have concluded that the hot core G24.78+0.08
contains at least 4 high-mass YSOs with different ages: more precisely, $t_{\rm
B} > t_{\rm A} > t_{\rm C} > t_{\rm D}$. In the light of our results, we can
now infer that the embedded YSOs could be 5, being A1 probably older than A2.
In addition, the fact that source C is colder ($T\simeq30$~K; see Codella et al.~\cite{codella97}) than source A2 ($T\simeq127$~K; see Sect.~\ref{g24_temp}) suggests that A2 is probably older than C. Thus, the
ages of the YSOs in G24.78+0.08 seem to be in the order $t_{\rm B} > t_{\rm A1} >
t_{\rm A2} > t_{\rm C} > t_{\rm D}$.

\subsubsection{Velocity field: the rotating toroid}
\label{models_g24}

\begin{figure}
\centerline{\includegraphics[angle=0,width=8.5cm]{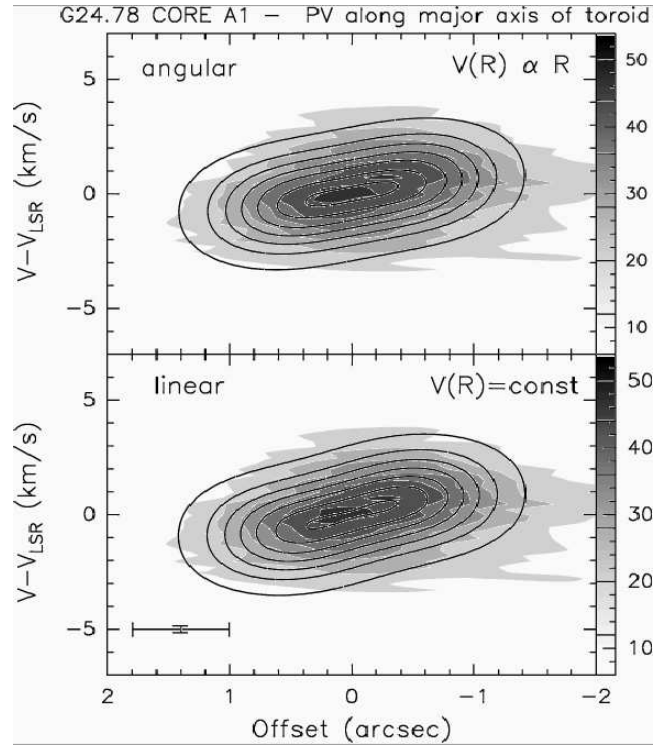}}
\caption{Same as Fig.~\ref{g31_pv_models} along the major axis of the toroid (P.A.\ $\sim 50\degr$) in core A1 in G24.78+0.08. The systemic velocity of  G24.78+0.08 is 111~\kms. The contour levels are 4, 12, 20, 28, 36, 44, and 50~\jy.}
\label{g24_pv_models_a1}
\end{figure}

In Paper~I we report on the detection of a clear velocity gradient in both core
A1 and A2, with \Vlsr\ increasing steadily along a direction at a P.A.\ $\sim 50\degr$, which is perpendicular to the axis of the molecular outflow detected toward
core A by Furuya et al.~(\cite{furuya02}). As already discussed in Paper~I, the
most plausible interpretation for such velocity gradients is that the cores
have toroidal structures undergoing rotation about the corresponding outflow
axis. In such a scenario, the two cores A1 and A2 would correspond to two
distinct rotating toroids. However, the separation of the cores is small, 
$1\farcs5$ or 12000~AU, and we cannot rule out the possibility that they are
part of the same geometrically thick rotating toroid, or pipe, elongated along
the direction of the outflow. In such a scenario, assuming core A1 as the
powering source of the outflow, the velocity gradient detected in the region
could be tracing a general rotation of this pipe, related to the rotation
associated with the launching of the jet. Taking into 
account jet models that predict that the flow moves along the magnetic 
surfaces conserving its angular momentum after the Alfv\'en surface 
(e.g.\ Spruit~\cite{spruit96}; K\"onigl \& Pudritz~\cite{konigl00}), 
the jet launching radius should be about half of the radius of the toroid 
in A1 in order to reach the maximum velocity in A2, which is measured 
at a distance of $\sim 0.02$~pc. Such a launching radius is consistent 
with the data, and thus, it calls for further higher angular resolution observations aimed to face the open question.

\begin{figure}
\centerline{\includegraphics[angle=-90,width=7.5cm]{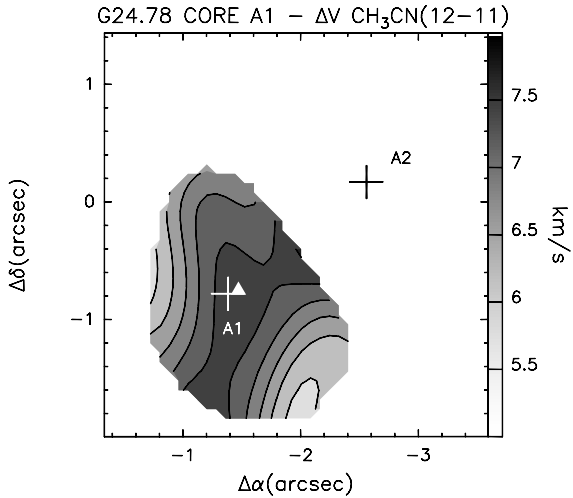}}
\caption{Line width plot of the \MCN\ (\jdo) toward core A1 in G24.78+0.08. The contour levels range from 5.5 to 7.5~\kms\ in steps of 0.5~\kms. The crosses mark the positions of the 1.4~mm continuum emission peaks.  The white triangle marks the position of the 1.3~cm continuum emission peak from the \HII region (Codella et al.~\cite{codella97}).}
\label{g24_linewidth}
\end{figure}

Due to the difficulty to establish which of the two cores, A1 or A2,  is
powering the bipolar outflow, we arbitrarily assume that it is core A1, as also done in Paper~I, because its association with an UC \HII region indicates that a YSO has already formed. Thus,
we model only the rotating structure associated with core~A1. Note that the PV plots for both cores are very similar, and thus, the results and conclusions derived from the modeling, with the exception of the exact value of the model parameters, apply also to core~A2. The
velocity shift measured over an extent of $\sim 9000$~AU is $\sim 3$~\kms, that is, a
velocity gradient of about 70~\kms\ pc$^{-1}$. 
The kinematics of the gas toward core A1 can be investigated with a PV cut of the \MCN\ (\jdo) $K=3$ emission along a direction (P.A. $\sim
50\degr$) roughly perpendicular to the outflow axis
(Fig.~\ref{g24_pv_models_a1}). As done for G31.41+0.31, we modeled the
emission by assuming power-law distributions for the temperature,
and the velocity field in the toroid (see Sect.~\ref{g31-models}), and adopted constant density in the core. The models computed were with constant rotation velocity and with constant angular
velocity, because Keplerian rotation is not possible due to the huge mass of the toroid. For the temperature distribution we again adopted a temperature power law $T\propto R^{-3/4}$.  As can be seen in Fig.~\ref{g24_pv_models_a1}, from the
modeling itself it is not possible to distinguish between constant angular velocity ($v \propto R$; top panel) or constant rotation ($v =$ constant; bottom panel), because both synthetic PV diagrams fit the data reasonably
well.  We computed models with
and without temperature gradients, and both fitted the data equally well. The fact that
a molecular outflow and an UC \HII region have been detected in association with
this core implies the existence of an embedded source powering it.
Therefore, although not clearly seen in our data, one may reasonably assume the existence of a temperature gradient driven by the central heating source.
Thus, we adopted a temperature distribution $T\propto R^{-3/4}$ for the 
models. The fact that the G24.78+0.08 \MCN\ emission, unlike that of 
G31.41+0.31, has not
got a ``hole" at the center, is consistent with the gas emission of core A1 in
G24.78+0.08 being optically thinner than that of G31.41+0.31.

The best fit model parameters are given in Table~\ref{table_models}. The value
of $R_{\rm inn}$ is $0\farcs$3 ($\sim 2300$~AU), which is the size of the UC
\HII region at 1.3~cm at the $3\sigma$ emission level as recently observed with
the VLA in the A configuration (Beltr\'an, private communication). The value of
$v_{\rm rot}$ is consistent with that derived directly from the velocity
gradient in Paper~I. The $R_{\rm out}$ derived from the model is roughly twice
the value measured from the 1.4~mm continuum emission at 50\% of the peak (see
Table~1 in Paper~I). The value of $N_{\rm CH_3CN}^{\rm peak}$ is $\sim 5$ times
lower than the column density derived by means of the RD method (see
Sect.~\ref{g24_temp}). As already mentioned for G31.41+0.31, this is due to the fact that the models do not take into account the
clumpiness of the region. 

\begin{figure}
\centerline{\includegraphics[angle=0,width=9cm]{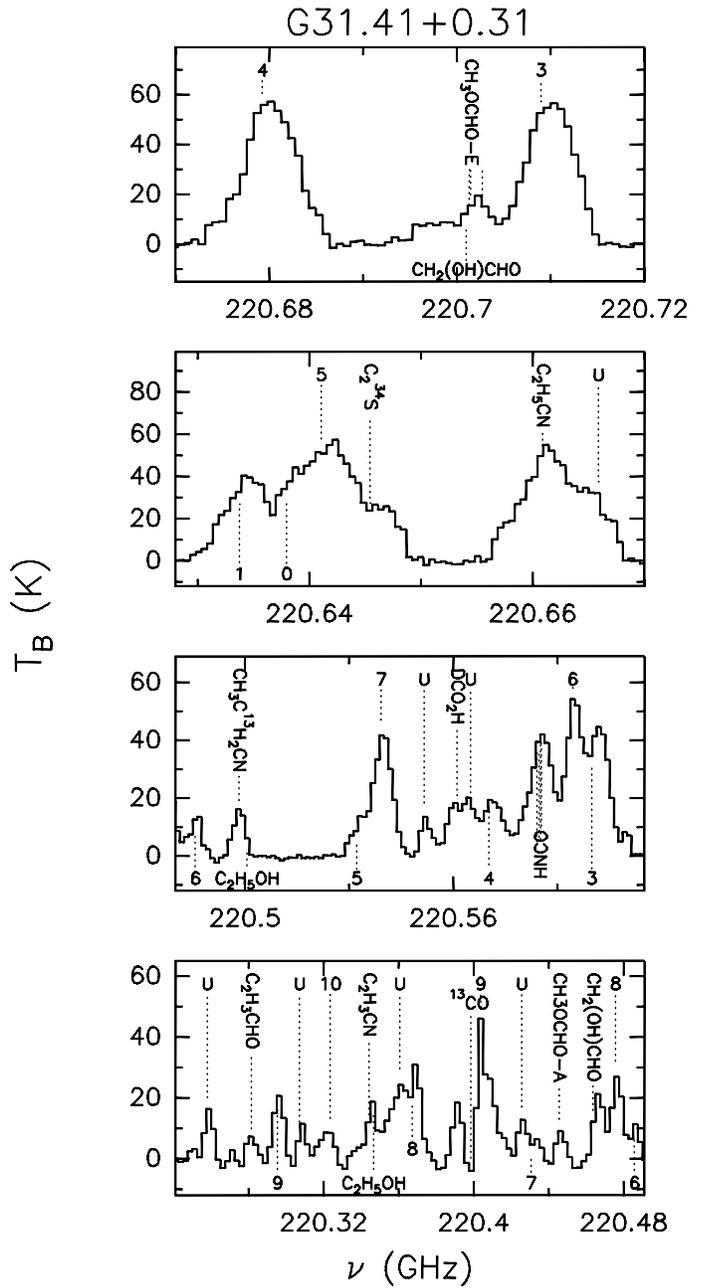}}
\vspace{0.5cm}
\caption{Serendipitous detections toward the center of G31.41+0.31 obtained with the PdBI. 
The vertical labels along the profiles point out the identified lines, reported in Tables \ref{table_sugar_detec} and \ref{table_blended}, and
the unidentified peaks, listed in Table~\ref{table_blended}.
The $y$-axis scale is expressed in brightness temperature. 
The dotted vertical lines are corresponding numbers indicate the position 
of the \MCN\ (\jdo) $K$-components in the upper part part of each spectra, 
and of the \MCNI\ (\jdo) $K$-components in the lower part part.}
\label{g31-sugar}
\end{figure}

\begin{figure}
\centerline{\includegraphics[angle=0,width=9cm]{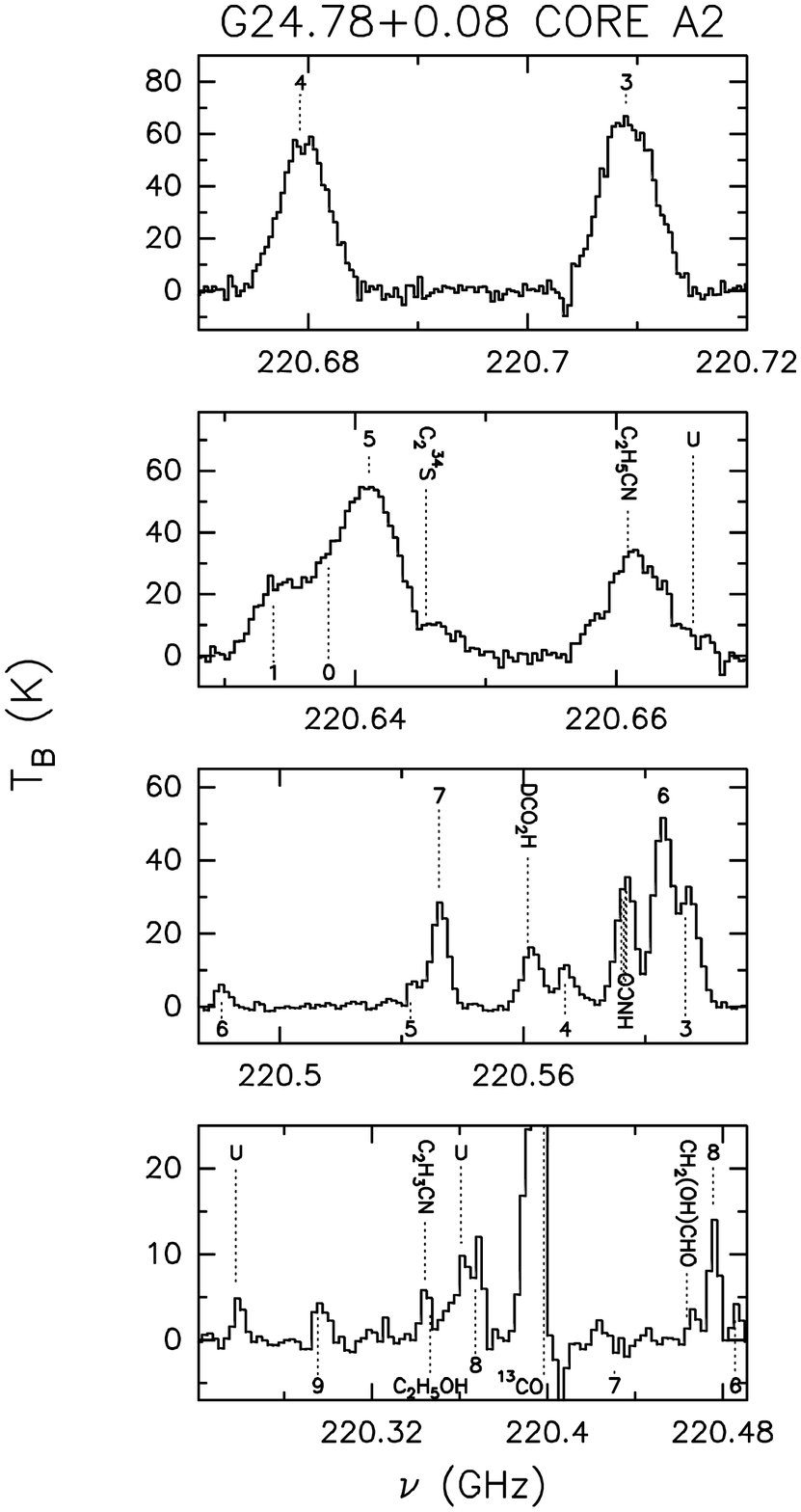}}
\vspace{0.5cm}
\caption{Serendipitous detections toward the peak position of core A2 in G24.78+0.08 obtained with the PdBI. 
The vertical labels along the profiles point out the identified lines, reported in Tables \ref{table_sugar_detec} and \ref{table_blended}, and
the unidentified peaks, listed in Table~\ref{table_blended}.
The $y$-axis scale is expressed in brightness temperature. 
The dotted vertical lines are corresponding numbers indicate the position 
of the \MCN\ (\jdo) $K$-components in the upper part part of each spectra, 
and of the \MCNI\ (\jdo) $K$-components in the lower part part.}
\label{g24-sugar_spectra}
\end{figure}

The dynamical mass computed with these values assuming that the toroid is seen
edge-on by the observer is $M_{\rm dyn} \sim 34~M_{\odot}$. This value is much
smaller than the mass of the core, and, as for G31.41+0.31,  suggests that such
a structure is unstable against gravitational collapse. The infall of material
onto the embedded millimeter source is also suggested by the line widths,
$\Delta V$, measured in the toroid (see Fig.~\ref{g24_linewidth}). Note that
for G24.78+0.08 the line widths have been measured from \MCN, because it has
not been possible to properly measure them from \MCNII, probably due to an
opacity effect. As seen in the figure, $\Delta V$ steadily increases toward the
center of the toroid and peaks at the position of the millimeter continuum
source. The estimate of $v_{\rm inf}$ derived from the difference between
$\Delta V$ at the position of the 1.4~mm continuum source and that at the edge
of the toroid (see Sect.~\ref{g31-models}) is $v_{\rm inf} \simeq 1.2$~\kms\ in
this case. This value is slightly lower than the $v_{\rm rot}$ obtained from
the models or from the velocity gradient in the toroid. The accretion rate
derived for such a $v_{\rm inf}$ is $\dot M_{\rm acc}\sim
9\times10^{-3}~M_\odot$~yr$^{-1}$. Such a huge accretion rate supports the
disk-accretion scenario also for the formation of the massive YSO embedded in
core A1 in G24.78+0.08.

\begin{table*}
\caption[] {Serendipitous detections toward G31.41+0.31 and G24.78+0.08 (A1 and A2)$^{\rm a}$}
\label{table_sugar_detec}
\begin{tabular}{llrccccc}
\hline
\multicolumn{1}{c}{Molecule} &
\multicolumn{1}{c}{Transition} &
\multicolumn{1}{c}{Rest frequency} &
\multicolumn{1}{c}{$E_{\rm u}$} &
\multicolumn{1}{c}{\Vlsr} &
\multicolumn{1}{c}{FWHM} &
\multicolumn{1}{c}{$T_{\rm B}$} &
\multicolumn{1}{c}{$\int{T_{\rm B}\,{\rm d}V}$} \\
\multicolumn{1}{c}{} &
\multicolumn{1}{c}{} &
\multicolumn{1}{c}{(MHz)} &
\multicolumn{1}{c}{(K)} &
\multicolumn{1}{c}{(km s$^{-1}$)} &
\multicolumn{1}{c}{(km s$^{-1}$)} &
\multicolumn{1}{c}{(K)} &
\multicolumn{1}{c}{(K km s$^{-1}$)} \\
\hline
\multicolumn{8}{c}{G31.41+0.31} \\
\hline
C$_2$H$_5$CN$^{\rm b}$ &(25$_{2,24}$--24$_{2,23}$) & 220660.92~\, & 143  &$96.3\pm0.1$ &$7.7\pm0.2$ &$50.6\pm3.3$ &$415\pm12$ \\
C$_2$$^{34}$S$^{\rm c}$ & (24$_{23}$--23$_{23}$) & 220645.47~\, & 187 &$96.1\pm0.5$ &$7.2\pm0.5$ &$17.4\pm4.2$ &$133\pm15$ \\
HNCO$^{\rm d}$ &(10$_{1,9}$--9$_{1,8}$) &220585.20$^{\rm e}$ & 102 &$96.6\pm0.4$ &$12.2\pm1.2^{\rm e}$ &$39.7\pm4.3$ &$518\pm39$ \\
DCO$_2$H$^{\rm f}$ & (10$_{6,4}$--9$_{6,3}$) & 220561.16$^{\rm g}$ & 139 &$98.4\pm0.5$ &$6.4\pm1.5$ &$14.4\pm2.1$ &$~\,98\pm$44  \\
CH$_2$(OH)CHO$^{\rm h}$ &(20$_{2,18}$--19$_{3,17}$) & 220463.87~\, & 120 &$93.1\pm0.8^{\rm h}$ &$8.2\pm1.8$ &$22.1\pm3.7$ &$193\pm$36  \\
CH$_3$OCHO-A &(25$_{11,15}$--26$_{9,18}$) & 220445.79~\, & 272 &$96.0\pm1.1$ &$7.9\pm2.4$ &$10.9\pm3.4$ &$~\,92\pm$26 \\
\hline
\multicolumn{8}{c}{G24.78+0.08 core A1} \\
\hline
C$_2$H$_5$CN$^{\rm b}$ &(25$_{2,24}$--24$_{2,23}$) & 220660.92~\, & 143  &$111.4\pm0.1$ &$4.3\pm0.2$ &$24.3\pm2.0$ &$111\pm5$ \\
HNCO$^{\rm d}$ &(10$_{1,9}$--9$_{1,8}$) &220585.20$^{\rm e}$ & 102 &$111.8\pm1.7$ &$7.0\pm1.7^{\rm e}$ &$19.9\pm1.7$ &~$148\pm28$ \\
DCO$_2$H$^{\rm f}$ & (10$_{6,4}$--9$_{6,3}$) & 220561.16$^{\rm g}$ & 139 &$111.7\pm0.2$ &$3.9\pm0.7$ &$8.9\pm1.1$ &$~\,37\pm$5  \\
CH$_2$(OH)CHO$^{\rm h}$ &(20$_{2,18}$--19$_{3,17}$) & 220463.87~\, & 120 &---$^{\rm i}$ &---$^{\rm i}$ &$3.0\pm0.5^{\rm i}$ &$~\,41\pm3^{\rm i}$  \\
\hline
\multicolumn{8}{c}{G24.78+0.08 core A2} \\
\hline
C$_2$H$_5$CN$^{\rm b}$ &(25$_{2,24}$--24$_{2,23}$) & 220660.92~\, & 143  &$110.4\pm0.1$ &$6.7\pm0.4$ &$33.4\pm1.4$ &$237\pm11$ \\
C$_2$$^{34}$S$^{\rm c}$ & (24$_{23}$--23$_{23}$) & 220645.47~\, & 187 &$110.5\pm0.6$ &$6.1\pm1.0$ &$10.7\pm1.6$ &$70\pm13$ \\
HNCO$^{\rm d}$ &(10$_{1,9}$--9$_{1,8}$) &220585.20$^{\rm e}$ & 102 &$110.9\pm0.1$ &$7.5\pm1.2^{\rm e}$ &$35.3\pm0.8$ &$281\pm5$ \\
DCO$_2$H$^{\rm f}$ & (10$_{6,4}$--9$_{6,3}$) & 220561.16$^{\rm g}$ & 139 &$109.7\pm0.2$ &$7.2\pm0.4$ &$16.1\pm1.1$ &$124\pm$7  \\
CH$_2$(OH)CHO$^{\rm h}$ &(20$_{2,18}$--19$_{3,17}$) & 220463.87~\, & 120 &$107.1\pm0.3^{\rm i}$ &$6.4\pm0.7$ &$3.9\pm0.4$ &$26\pm$3  \\
\hline
\end{tabular}

(a) Spectra taken toward the peak positions of G31.41+0.31 and cores A1 and A2 in G24.78+0.08. \\
(b) Blended with the 220665.91 unidentified line (see Table~\ref{table_unidentified}). \\
(c) Blended with the $K$=5 component of \MCN\ (\jdo). \\
(d) Blended with the $K$=6 and $K$=4 components of \MCN\ and \MCNI\ (\jdo), respectively. \\ 
(e) The HNCO (10$_{1,9}$--9$_{1,8}$) line splits into six hyperfine components with $\Delta$$F$=0,$\pm$1 spread
over 0.6 MHz and thus not resolved with the present spectral resolution. The frequency refers to the 
$F$--$F^{'}$=10--9 line.\\
(f) This line could be blended with the high excitation ($E_{\rm u}$ = 625 K)  C$_2$H$_3$CN (10$_{3,7}$--10$_{2,8}$)
line at 220651.39 MHz. \\
(g) The rest frequency corresponds also to the DCO$_2$H (10$_{6,5}$--9$_{6,4}$) line, which has the same excitation. \\
(h) According to the databases for molecular spectroscopy, glycolaldehyde results to be the most probable 
candidate to identify the present emission at this frequency. If confirmed, this would be the first detection of CH$_2$(OH)CHO
outside the Galactic center (see text). \\
(i) Bad fit because of non-Gaussian profile.  
\end{table*}

\section{Other molecules in G31.41+0.31 and G24.78+0.08}

Figures~\ref{g31-sugar} and \ref{g24-sugar_spectra} show some portions of the
spectra obtained toward G31.41+0.31 and core A2 in G24.78+0.08, respectively.
Taking into account that the spectra for both core A1 and A2 in G24.78+0.08 are
quite similar, we have only plotted the spectra toward A2, where the detections
are more clearly visible. In these figures one can see that besides the
expected CH$_3$CN and $^{13}$CH$_3$CN patterns, numerous other lines have been 
serendipitously detected. These lines have been identified, when possible, by
using the Jet Propulsion Laboratory, Cologne, and Lovas databases for molecular
spectroscopy. Emission due to C$_2$H$_5$CN, C$_2$$^{34}$S, HNCO, DCO$_2$H, and
CH$_3$OCHO-A high excitation (100--270 K) transitions have been measured. For
these lines, Table~\ref{table_sugar_detec} lists the transition,  the rest
frequency, the upper level excitation $E_{\rm u}$ (K), and line parameters
obtained with a Gaussian fit, i.e.\ the velocity $V_{\rm LSR}$  (km s$^{-1}$),
the FWHM line width (km s$^{-1}$), the peak brightness  temperature $T_{\rm B}$
(K), and the integral under the line $\int$$T_{\rm B}$$dV$ (K km s$^{-1}$).

\begin{figure*}
\centerline{\includegraphics[angle=-90,width=12.5cm]{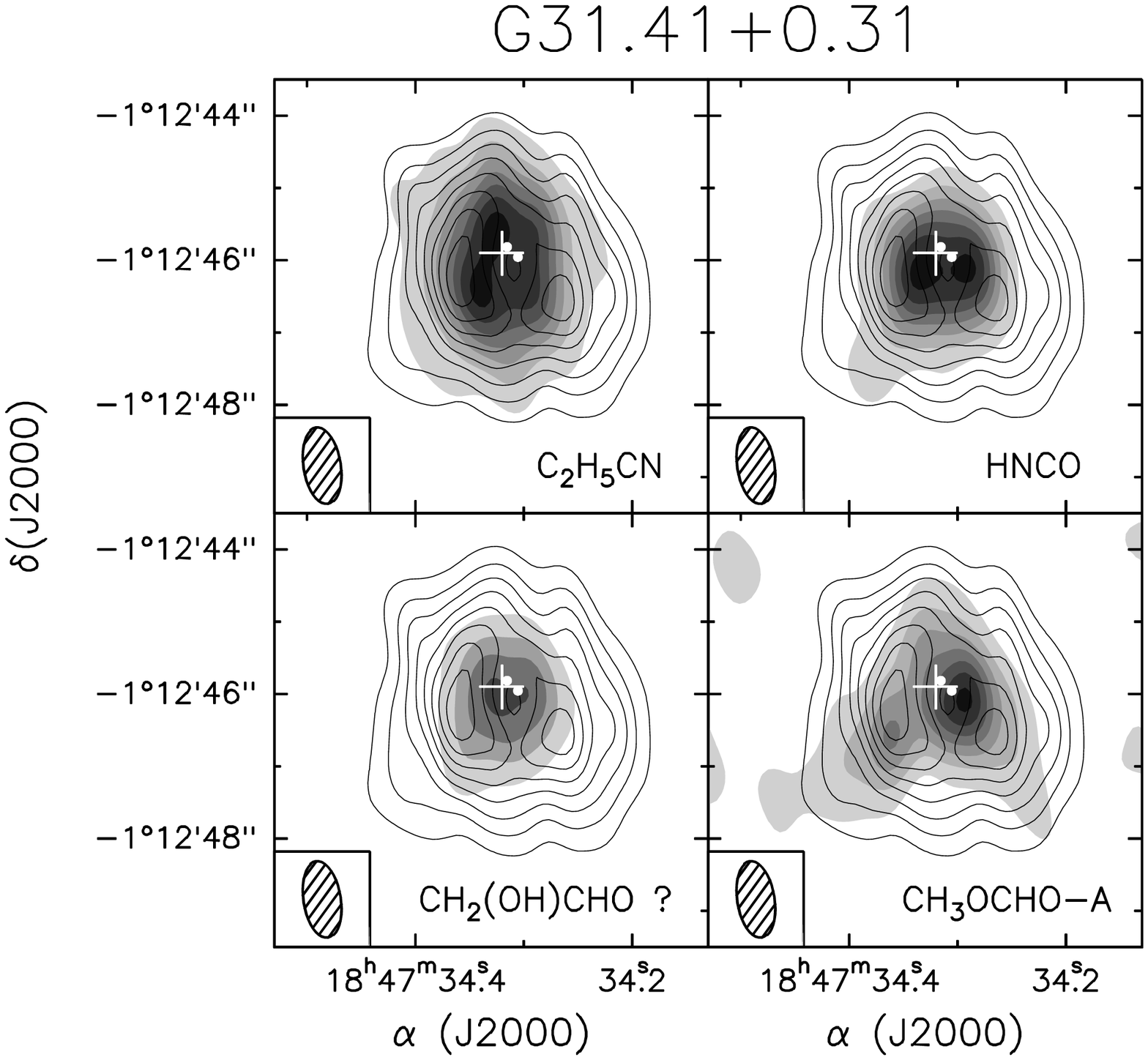}}
\caption{Overlay of the PdBI \MCN\ (\jdo) emission averaged under the $K=0$, 1, and 2 components ({\it contours}) and the C$_2$H$_5$CN (25$_{2,24}$--24$_{2,23}$) emission ({\it top left}), the HNCO (10$_{1,9}$--9$_{1,8}$) emission ({\it top right}), the CH$_2$(OH)CHO (20$_{2,18}$--19$_{3,17}$) emission ({\it bottom left}), and the CH$_3$OCHO-A (25$_{11,15}$--26$_{9,18}$) emission ({\it bottom right}) toward G31.41+0.31. Contours are the same as in Fig.~\ref{g31_ch3cn_1mm}. Greyscale levels range 
from 0.10 to 1.18\jy\ in steps of 0.18\jy\ for C$_2$H$_5$CN, from 0.10 to 0.94\jy\ in steps of 0.14\jy\ for HNCO, from 0.10 to 0.46\jy\ in steps of 0.12\jy\ for CH$_2$(OH)CHO, and from 0.01 to 0.13\jy\ in steps of 0.02\jy\ for CH$_3$OCHO-A. The synthesized beam is shown in the lower left-hand corner. The white cross marks the position of the 1.4~mm continuum emission peak, and the white dots the 7~mm continuum emission peaks detected by Hofner (private communication).}
\label{g31_molecules}
\end{figure*}

\begin{figure*}
\centerline{\includegraphics[angle=-90,width=12.5cm]{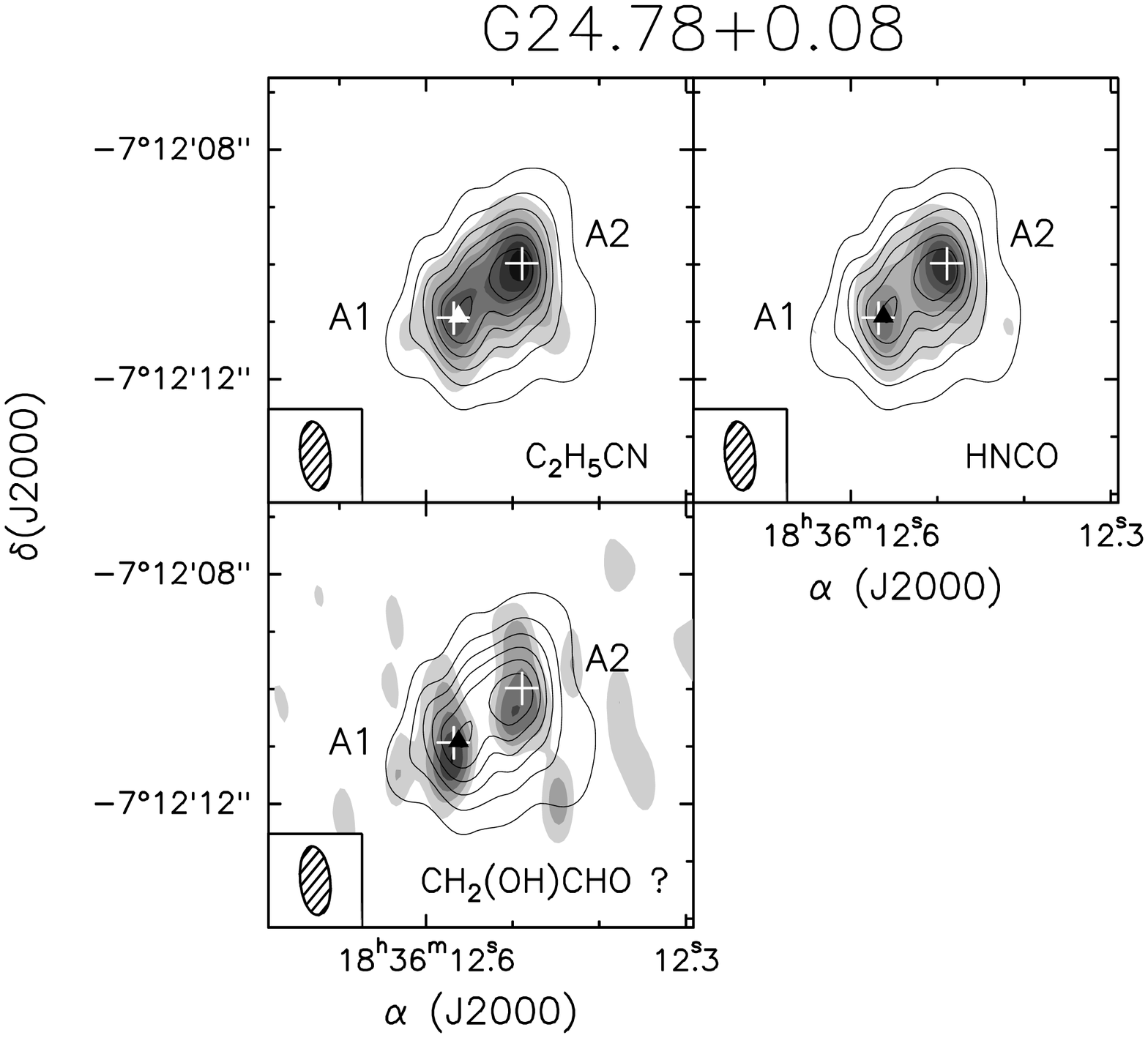}}
\caption{Overlay of the PdBI \MCN\ (\jdo) emission averaged under the $K=0$, 1, and 2 components ({\it contours}) and the C$_2$H$_5$CN (25$_{2,24}$--24$_{2,23}$) emission ({\it top left}), the HNCO (10$_{1,9}$--9$_{1,8}$) emission ({\it top right}), and the CH$_2$(OH)CHO (20$_{2,18}$--19$_{3,17}$) emission ({\it bottom left}) toward cores A1 and A2 in G24.78+0.08. Contours levels range from 0.10 to 1.00\jy\ in steps of 0.18\jy. Greyscale levels range 
from 0.10 to 0.70\jy\ in steps of 0.10\jy\ for C$_2$H$_5$CN, from 0.10 to 0.60\jy\ in steps of 0.10\jy\ for HNCO, and from 0.02 to 0.10\jy\ in steps of 0.02\jy\ for CH$_2$(OH)CHO. The synthesized beam is shown in the lower left-hand corner. The white crosses denote the position of the 1.4~mm continuum emission peaks. The triangle marks the position of the VLA 1.3~cm continuum emission peak (Codella et al.~\cite{codella97}).}
\label{g24-molecules}
\end{figure*}

\begin{table*}
\caption[] {Other molecular tracers detected toward G31.41+0.31 and G24.78+0.08}
\label{table_blended}
\begin{tabular}{ccclcc}
\hline
\multicolumn{1}{c}{Obs. frequency$^{\rm a}$} &
\multicolumn{1}{c}{$d\nu$} &
\multicolumn{1}{c}{Source} &
\multicolumn{3}{c}{Line candidates} \\
\multicolumn{1}{c}{(MHz)} &
\multicolumn{1}{c}{(MHz)} &&
\multicolumn{1}{c}{Transition} &
\multicolumn{1}{c}{Frequency(MHz)} &
\multicolumn{1}{c}{$E_{\rm u}$(K)} \\
\hline
\multicolumn{6}{c}{Blended lines} \\
\hline
$220702.17\pm0.15^{\rm b}$ & 0.7  &G31 & CH$_3$OCHO-E (10$_{4,7}$--9$_{3,7}$) & 220702.75 &43 \\
&    & & CH$_3$OCHO-E (24$_{3,22}$--24$_{2,23}$) & 220701.52 &169 \\
&    & & CH$_3$OCHO-E (24$_{3,22}$--24$_{1,24}$) & 220701.36 &169 \\
&    & & CH$_2$(OH)CHO (40$_{9,32}$--40$_{8,33}$) & 220701.03 &511 \\
\hline
$~220498.34\pm0.08^{\rm c}$ & 1.3 &G31 & C$_2$H$_5$OH (19$_{5,14}$--19$_{4,15}$) & 220500.70 &250 \\
&    & & CH$_3^{13}$CH$_2$CN (7$_{4,3}$--6$_{3,4}$) & 220498.40 &29 \\
\hline
$220345.51\pm0.16$& 2.5 &G31, G24 & C$_2$H$_5$OH (7$_{3,4}$--7$_{1,7}$) & 220346.86 &91 \\
&   &  & C$_2$H$_3$CN (9$_{4,5}$--10$_{3,8}$) & 220344.45 &55 \\
\hline
\end{tabular}

(a) The observed frequencies have been measured from the spectra toward G31.41+0.31 due to a better signal-to-noise ratio of the profiles. \\  
(b) Blended with a non Gaussian profile, which produces a plateau in the observed profiles (see Fig.~\ref{g31-sugar}). A possible candidate for the emission is C$_2$H$_3$CN$^{\rm b}$ (10$_{3,7}$--10$_{2,8}$) at a frequency of 220695.42~MHz. \\
(c) This line could be blended with the very high excitation ($E_{\rm u}$ = 1005 K) CH$_3$OCHO-E (46$_{0,46}$--45$_{4,41}$)
line at 220667.14 MHz. \\

\end{table*}

A special case is represented by a peak at 220.46 GHz (see
Table~\ref{table_sugar_detec}), where the only reasonable candidate to our
knowledge is the transition  $J_{\rm K_-K_+}$=20$_{\rm 2,18}$--19$_{\rm 3,17}$
of glycolaldehyde (CH$_2$(OH)CHO), whose excitation (120 K) well fits with the
temperature measured toward  G31.41+0.31 and G24.78+0.08. Glycolaldehyde is an
isomer to methyl formate (CH$_3$OCHO), here detected at 220.45 GHz. Whereas the
detection of CH$_3$OCHO is  not surprising since it is a common species in  hot
cores (e.g.\ van Dishoeck \& Blake~\cite{dishoeck98}), CH$_2$(OH)CHO has been
detected up to date only toward  Sagittarius B2N (Hollis et
al.~\cite{hollis00}). The reasons could be that complex interstellar molecules
seem to prefer the C--O--C backbones with respect to the C--C--O ones (see
Fig.~2 of Hollis et al.~\cite{hollis00}).  Note that other CH$_2$(OH)CHO lines
lie inside the spectral windows here investigated, but unfortunately they are
closely blended with other bright lines. In conclusion, we can only claim a
tentative detection of CH$_2$(OH)CHO toward  G31.41+0.31 and G24.78+0.08, which
calls for further observations to confirm  the occurrence of glycolaldehyde
emission in hot cores. Note that there is a difference between the line rest frequency and the frequency of the observed peak ($\sim$2 MHz), as indicated by a slightly different peak velocity from the \Vlsr\ of the region (Table~\ref{table_sugar_detec}). If the detection of glycolaldehyde is
confirmed, this would  imply that the rest frequency listed in the spectral catalogues should be refined.

Table~\ref{table_blended} reports the observed frequency, the corresponding spectral resolution and the line candidates with their frequency and
excitation for those spectral features that could not be unambiguously identified. The candidate transitions have frequencies that fall inside the spectral features, and were selected on the basis of the following criteria: (i) emission lines known to be hot core tracers, such as
CH$_3$OCHO-E, C$_2$H$_3$CN, and C$_2$H$_5$OH (e.g.\ van Dishoeck \&
Blake~\cite{dishoeck98}; Cazaux et al.~\cite{cazaux03}), and (ii) lines 
with excitations
in agreement with the temperatures measured toward G31.41+0.31 and
G24.78+0.08 ($E_{\rm u}\lesssim 500$~K). Finally, Table~\ref{table_unidentified} reports the peak
frequencies and the relative spectral resolution of the unidentified lines
toward G31.41+0.31 and G24.78+0.08. In conclusion, the gas associated with G31.41+0.31 results
more chemically rich than that in G24.78+0.08: this could be
caused by a different evolutionary stage or it could simply be a luminosity
effect being G31.41+0.31 more luminous than G24.78+0.08.

\begin{table}
\caption[] {Unidentified lines detected toward G31.41+0.31 and G24.78+0.08}
\label{table_unidentified}
\begin{tabular}{ccc}
\hline
\multicolumn{1}{c}{Frequency$^{\rm a}$} &
\multicolumn{1}{c}{Spectral resolution} &
\multicolumn{1}{c}{Source} \\
\multicolumn{1}{c}{(MHz)} &
\multicolumn{1}{c}{(MHz)}  \\
\hline
$220665.91\pm0.15$ & 2.5 &G31, G24  \\
$220564.07\pm0.29$ & 1.3 &G31 \\
$220551.97\pm0.41$ & 1.3 &G31 \\
$220426.40\pm0.18$ & 2.5 &G31  \\
$220361.37\pm0.50$ & 2.5 &G31, G24 \\
$220308.68\pm0.43$ & 2.5 &G31 \\
$220258.91\pm0.31$ & 2.5 &G31 \\
\hline
\end{tabular}

(a) The frequencies have been measured from the spectra toward G31.41+0.31, due to a better signal-to-noise ratio of the profiles, assuming a systemic velocity \Vlsr = 97~\kms. \\

\end{table}

In order to understand which regions are traced by these complex molecules, the
maps of the emission due to the unblended lines is shown in
Figs.~\ref{g31_molecules} (for G31.41+0.31) and \ref{g24-molecules} (for G24.78+0.08).  The distributions of C$_2$H$_5$CN, HNCO, CH$_3$OCHO-A, and
CH$_2$(OH)CHO (greyscale) have been compared with that of the $K$=0, 1, and 2
components of CH$_3$CN (contours). For G31.41+0.31, it is possible
to see that the emission due to these molecules comes from the central region
where the protostars are located as indicated by the positions of the 1.4~mm
(white cross) and 7~mm (white dots) peaks. This result confirms the presence
of a hot molecular core associated   with G31.41+0.31, in agreement with the
increase of temperature and density toward the center of the region as 
measured with CH$_3$CN (see Fig.~\ref{g31_trot_ntot}).   Moreover,
Fig.~\ref{g31_molecules} shows that the emission from different tracers peaks at slightly
different positions, indicating that C$_2$H$_5$CN, HNCO, CH$_3$OCHO-A, and
CH$_2$(OH)CHO are tracing different portions of the hot core.
Figure~\ref{g24-molecules} shows a similar scenario for G24.78+0.08: the 
C$_2$H$_5$CN, HNCO, and CH$_2$(OH)CHO (CH$_3$OCHO-A has not been detected
toward G24.78+0.08) emission traces the two hot cores, A1 and A2, confirming that each of them is associated with a distinct site of star formation and is hosting YSOs.

\section{Conclusions}

We have analyzed the millimeter high angular resolution data,
obtained with the BIMA and IRAM PdBI interferometers, of the dust and gas
emission toward the high-mass star forming regions G31.41+0.31 and G24.78+0.08.
The aim was a thorough study of the structure and kinematics of the rotating
toroids detected in both hot cores in Paper~I.

The \MCN\ (\jdo) emission of the toroids in G31.41+0.31 and core A1 in
G24.78+0.08 has been modeled assuming that it arises from a disk-like structure
seen edge-on, with a radial velocity field. For G31.41+0.31 the model properly
fits the data for a velocity $v_{\rm rot}\simeq 1.7$~\kms\ at the outer radius
$R_{\rm out}\simeq 13400$~AU and an inner radius $R_{\rm inn}\simeq 1340$~AU,
while for core A1 in G24.78+0.08 the best fit is obtained for $v_{\rm
rot}\simeq 2.0$~\kms\ at $R_{\rm out}\simeq 7700$~AU and $R_{\rm inn}\simeq
2300$~AU. Unlike the Keplerian disks detected around less luminous stars such
as IRAS~20126+4104 (e.g.\ Cesaroni et al.~\cite{cesa97}), NGC~7538S (Sandell et
al.~\cite{sandell03}), or M17 (Chini et al.~\cite{chini04}), these toroids are
not undergoing Keplerian rotation. From the modeling itself, however, it is not
possible to distinguish between constant rotation or constant angular velocity,
since both velocity fields suitably fit the data. The best fit models have
been computed assuming for the toroids a temperature gradient of the type $T
\propto R^{-3/4}$, typical of accretion circumstellar disks
(Natta~\cite{natta00}), with a temperature at the outer radius $T_{\rm
out}\simeq 100$~K for both cores.

The $M_{\rm dyn}$ needed for equilibrium derived from the
models is much smaller than the mass of the cores. This suggests, as
already pointed out in Paper~I, that such toroids are unstable and
undergoing gravitational collapse. The collapse is supported also by the fact that the \MCNII\ or \MCN\ line width measured in the cores increases toward the center of the
toroids in G31.41+0.31 and G24.78+0.08 A1. The estimates of
$v_{\rm inf}$ and $\dot M_{\rm acc}$ are $\sim 2$~\kms\ and $\sim
3\times10^{-2}~M_\odot$~yr$^{-1}$ for G31.41+0.31, and $\sim 1.2$~\kms\
and $\sim 9\times10^{-3}~M_\odot$~yr$^{-1}$ for G24.78+0.08 A1.
Accretion rates that large could weaken the effect of stellar winds and radiation pressure thus allowing further accretion on the star. This would support theories according to which high-mass stars form like their lower mass counterparts through non-spherical accretion with large
accretion rates.

The values of $T_{\rm rot}$ and $N_{\rm CH_3CN}$, derived by means of the RD
method, for both G31.41+0.31 and the sum of cores A1 and A2 (core A of Codella et
al.~\cite{codella97}) in G24.78+0.08 are in the range 132--164~K and 2--$8\times10^{16}$~cm$^{-2}$.

For G31.41+0.31, the most plausible explanation for the apparent toroidal
morphology seen in the lower $K$ transitions of \MCN\ (\jdo) in Paper~I 
is self-absorption, which is caused by the high optical depth and
the existence of a temperature gradient in the core. Indeed, the $N_{\rm CH_3CN}$ presents a peak toward the apparent dip in the CH$_3$CN brightness temperature map.

\begin{acknowledgements}
We would like to thank Francesca Bacciotti, Daniele Galli and Malcolm Walmsley for helpful discussions and suggestions. 
\end{acknowledgements}


\end{document}